\documentstyle[12pt]{article}
\begin{document}

\centerline{\LARGE \bf Cascades initiated by EHE photons in}   
\centerline{\LARGE \bf the magnetic field of the Earth and the Sun}  

\begin{center}
{\bf W. Bednarek}\\
{\it Department of Experimental Physics, University of \L \'od\'z,\\
90-236 \L \'od\'z, ul Pomorska 149/153, Poland\\
bednar@krysia.uni.lodz.pl}
\end{center}

\begin{center}
{\large \bf Abstract\\}
\end{center}

The content of extremely high energy (EHE) photons in the highest energy
cosmic rays can be investigated by analyzing showers arriving 
from directions where the perpendicular component
of the Earth's magnetic field is high or showers arriving from 
the region surrounding the Sun.
We perform Monte Carlo simulations of cascades initiated by photons 
with parameters of the highest energy showers observed by the past and 
present detectors (AGASA, Fly's Eye, Yakutsk, Haverah Park). The purpose
is to  find out which events should cascade with high probability if initiated
by photons. It is shown that EHE photons
arriving from directions towards the magnetic poles
cascade with higher probability. Alternatively, the lowest probabilities  
of cascading are expected for photons arriving from directions of
the equator at the zenith angles equal to
the angle between location of the specific observatory and the
magnetic pole. 
We show that very unusual showers should arrive from direction of 
the Sun due to cascading of photons in the Sun's magnetosphere 
if EHE photons are numerous at the highest energies. The rate of such 
showers
is estimated on about one per ten years. However extraordinary lateral
distribution of secondary particles of such showers in the Earth's 
atmosphere may help to distinguish them from the ordinary showers arriving
isotropically from the sky.



\newpage

\section{Introduction.}

The spectrum of cosmic rays shows very interesting and 
unexpected features at the highest energies. Different experiments 
report a change in the spectral slope at $\sim 10^{19}$ eV which is
usually interpreted as a transit from Galactic to extragalactic
origin. This hypothesis is supported by the recent observations of
significant enhancements of cosmic rays from the direction of the 
galactic plane at energies $(0.8 - 2.0)\times 
10^{18}$eV~\cite{haetal98,bietal98} and the lack of anisotropy above
$10^{19}$eV~\cite{taetal99}. Such an explanation for the change in the
spectrum  at $\sim 10^{19}$ eV is also consistent with reports of the Fly'e Eye
group on the change of composition of cosmic rays from heavy to light at
energies above $10^{18}$ eV~\cite{gaetal93} (see also the conclusions 
based on
the AGASA data~\cite{haetal95} and their  re-analysis~\cite{dms98}). If the
highest energy cosmic rays are extragalactic,  then they  cannot travel from
cosmological distances because of their interaction with the microwave
background radiation (MBR). The propagation distances of heavy nuclei 
on fragmentation in the MBR have been recently re-calculated by
Stecker \& Salamon~\cite{ss98} and Epele \& Roulet~\cite{er98} 
(see the references therein for earlier calculations). Protons with 
energies
above $5\times  10^{19}$ eV produce charged and  neutral pions  in collisions
with the MBR  photons. The neutral pions decay in turn into
photons. Since the propagation distances of hadrons are relatively short (a
few tens of Mpc), it is  expected that the spectrum of cosmic rays above
$5\times 10^{19}$ eV should decline (so called the  Greisen-Zatsepin-Kuz'min
cut-off~\cite{gr66,zk66}). Such a cut-off in the spectrum is not found in the
recent AGASA data~\cite{taetal98}. However an interesting small deficit of
particles at around $10^{20}$eV is noted by Takeda~et al.~\cite{taetal98}. It
may suggest the emergence of a new component in the spectrum above $10^{20}$
eV.

EHE photons with energies $\sim 10^{20}$ eV are also absorbed in 
collisions with MBR on a distance scale of the order of several Mpc. 
This is comparable to the propagation distances of protons with 
such energies. However the propagation distances for photons 
above 
$10^{20}$ eV decreases but the propagation distances of protons 
do not change significantly. If the strong radio
background exists\cite{pb96}, the propagation distances for photons 
becomes longer than that ones for protons at energies greater than 
$\sim 10^{21}$ eV. The number of photons at energies above $10^{19}$ eV 
may be significant from two reasons. EHE cosmic ray protons, distributed  
uniformly
in the Universe, produce many photons in collisions with the MBR
photons~\cite{ww90,haetal90}. The EHE photons can be also efficiently 
produced
from decay of massive  particles (e.g. Higgs and gauge bosons), as
predicted by some more exotic theories (see~\cite{bs98} and  references
therein). Interestingly, this last model predicts that a new component 
in the  cosmic ray spectrum should emerge above $\sim 10^{20}$ eV, 
composed mainly from photons and neutrinos. 

Since the number of photons at the highest energy cosmic rays may be high,
it is reasonable to discuss the consequences of assumption that the
observed highest energy showers are due to the interaction of
photons. However photons with such energies should interact at first with 
the dipole magnetic field of the Earth. 
In this paper we compute the cascades initiated by  EHE photons using the 
parameters of the highest energy showers observed by the present and past
detectors. Moreover, we investigate the effects of cascading of EHE photons at
the locations of the  Southern and Northern Auger Observatories. 
The detailed investigation of the parameters of the showers observed by 
these detectors (the arrival directions, energies)
should give us information about the possible content of photons in the EHE
cosmic rays. 

The magnetic field 
of the Sun is about an order of magnitude stronger than that of the 
Earth. Therefore detection of secondary photons
from cascades initiated in the magnetic field of the Sun can  allow  
an investigation of the photon content in the EHE cosmic ray  
spectrum at energies about an order of magnitude lower, provided that 
a large enough detector of cosmic ray showers is available. 
We calculate the cascades
initiated by photons in the magnetic field of the Sun and discuss the
possibility of detection of extraordinary showers in the Earth's
atmosphere produced by
the  secondary photons from these cascades by the  Auger Observatories.

\section{Magnetic $e^\pm$ pair cascades.}  

An EHE photon with energy $E_\gamma$ can convert into an $e^\pm$ pair
in the magnetic field $B$ if the value of the parameter 
$\chi_\gamma = (E_\gamma/2mc^2)(B/B_{cr})$ 
(where $B_{\rm cr}\approx 4.414\times 10^{13}$G and $mc^2$ is the electron
rest mass) is high enough~\cite{er66}. 
The secondary $e^\pm$ 
pairs produce in turn synchrotron photons in the magnetic field.
For the electrons with the highest energies expected, the synchrotron 
photons are produced in the quantum domain and the efficiency of 
the process is determined by energy of the electron and the parameter 
$\chi_{\rm e} = (E_{\rm e}/mc^2)(B/B_{cr})$. The energies of  
synchrotron photons can be high enough to produce the  next generation 
of $e^\pm$ pairs.  We simulate the development of such a cascade
by using the Monte Carlo method and applying the exact 
quantum mechanical rates for $e^\pm$ pair
production by $\gamma$-ray photon and synchrotron emission by secondary
$e^\pm$ pairs as given by Baring~\cite{ba89}. In Fig.~1, we compare the
Baring's rates with the approximate Erber's rates~\cite{er66} for 
conversion of a photon into an $e^\pm$ pair and for emission of 
synchrotron
photons  by an electron (positron) by calculating the probabilities of
production  of $e^\pm$ pair with energy below $E_{\rm e}$ and 
the probabilities of emission of synchrotron photons with energies 
below
$E_{\rm syn}$. The top figures show the results for conversion of a photon 
into an $e^\pm$ pair and the bottom figures for emission of synchrotron 
photons by an electron. Significant differences between the Baring's 
and the Erber's rates are evident for large values  of $\chi_\gamma$ 
and $\chi_{\rm e}$. The Erber's rates give on  average lower energy 
$e^\pm$ pairs 
from conversion of photons and lower energy synchrotron photons from 
electrons. Therefore,
the cascades calculated with the Baring's rates are more penetrating.      
Note that for the surface magnetic fields typical for the Earth and the
Sun and for energies of events which may be detected by  the Auger
Observatory, the parameters $\chi$ can reach values above ten. 

In Fig.~\ref{fig1} we show the mean free path for conversion of photons
with energy $E_\gamma$ into $e^\pm$ pairs in a perpendicular magnetic field
with strength 0.3 G (full curve) and for production of synchrotron photons by
electrons (positrons)  with energy $E_e$ in this same magnetic field (dashed
curve). The mean free paths for other magnetic fields can be obtained 
from Fig.~\ref{fig1} by simple linear scaling of the present computations 
(shift to the left and down for stronger magnetic field). For
example, the mean free paths in the magnetic field of 3 G (a value typical
near the surface of the Sun) for photons with energies $10^{20}$ eV are
$\sim 7$ km and for the production  of synchrotron photons by electrons with
energy $10^{18}$ eV are  $\sim 0.38$ km. Photons
must have energies above  $\sim 3\times 10^{19}$ eV in order to cascade
efficiently in  the Earth's magnetic field with $B = 0.3$ G (the limit 
obtained from comparisons of the photon mean free path with the radius of  
the Earth). For the Sun this minimum photon energy is  $\sim 2\times 
10^{18}$ eV for $B = 3$ G.

\section{Photons cascading in the magnetic field of the Earth.}  

The magnetic field of the Earth is strong enough so that photons with
energies corresponding to energies estimated for the highest energy atmospheric
showers have chance to be converted in the magnetic field into $e^\pm$ 
pair. These 
effects of cascading of EHE photons have been already generally
discussed some time ago~\cite{ml81,ahetal91}. More recently general 
analysis of the interaction of EHE cosmic ray photons 
with energies $>10^{20}$ eV with the Earth's magnetosphere
has been performed by Karaku\l a et
al.~\cite{kmt96,kt95,kar97} and by others~\cite{sv97,ka97}. 
These authors use some simplifications (e.g. for the cross sections
of pair production and synchrotron emission, or neglect the 
inclination of the magnetic axis in respect to the Earth rotational 
axis). Specific showers are not analyzed in these papers under the 
hypothesis of their photonic origin. These authors do not discuss also 
what effects of cascading should we expect for the sites of both planned 
Auger Observatories.
In this section we analyze the cascades in the Earth's magnetosphere
initiated by photons with energies and
arrival directions of the showers observed by the Haverah Park,
Yakutsk, AGASA arrays and Fly's Eye detector. The results of this work 
will create an input for further analysis of cascades in the Earth's 
atmosphere under the hypothesis that these events are caused by photons.  
We expect that inclusion of effects of cascading in the Earth's
magnetosphere may have impact on the estimation 
of the primary energy of photons responsible for these events.
We predict the efficiency of cascading of photons
arriving to the locations of both Auger Observatories and localize 
the regions on the sky with the lowest and highest probability of 
cascading.

For the magnetic moment of the Earth we use the value $8\times 10^{25}$
G cm$^3$. The coordinates of the North Magnetic Pole are the following:
longitude 108$^{\rm o}$ and latitude 77.5$^{\rm o}$ (for the year 1980).

\subsection{Photons with energies of the observed EHE cosmic rays.}

We simulate the cascades initiated by photons with energies and 
arrival directions of the highest energy events observed by the air shower
detectors at different geographic locations, i.e. the Haverah Park event with
an energy of $1.59\times 10^{20}$ eV~\cite{hpcat}, the Yakutsk event with an
energy of $2.3\times 10^{20}$ eV~\cite{efetal91,afetal93}, the AGASA event with
an energy of $2.1\times  10^{20}$ eV~\cite{haetal94,taetal98}, and the Fly's
Eye event  with an energy of $3.2\times
10^{20}$ eV~\cite{bietal94}. The number of secondary 
$\gamma$-rays produced per  $\Delta(log  E_\gamma) = 0.1$ in these cascades
are shown in Fig.~\ref{fig2}. The spectra marked by the thick full histogram
are averaged over 100 primary photons and the spectra with the smallest 
and highest numbers of secondary photons, selected from these 100 
simulations,
are marked by the dotted and thin full histograms, respectively. It is evident
that fluctuations of the total number of secondary photons produced in
these cascades are large, of the order of $3 - 4$ around the average value. 
Photons with energies of
these highest energy events cascade with high probability, although photons
with the parameters of the AGASA event may not cascade at all with
the probability $\sim 3\%$. In all cases a significant part of energy of
primary photon is still carried by the secondary photons with energies above
$10^{19}$ eV.  On average, 6 photons produced in the case of primary photon
with the  parameters of the Fly's Eye event carry $\sim 31\%$ of total energy
of the event. In the case of other events these numbers are: the Haverah Park
event - 3.3 photons and $\sim 34\%$, the Yakutsk event - 3.9 photons and $\sim
25\%$, and the AGASA event - 4.6 photons and $\sim 41\%$. 

In Fig.~4 we show the probability ($\Delta N/\Delta log(R/R_{\rm Z})$)
of the first interaction of photons with parameters of these highest 
energy events as a function of distance from the Earth's surface $R$ 
in units of the radius of the Earth $R_{\rm Z}$. Additionally, we 
show such probability
for the supposed event with direction of the highest energy Fly's Eye
event but with energy $10^{21}$ eV. Photons with parameters of all 
observed events have the highest probability of first interaction
at the distance closer than one radius of the Earth. 
The event with energy $10^{21}$  eV will have the highest probability 
of interaction at the distance of $1 - 1.1$ radii of the Earth. 
Therefore, we conclude that our assumption on the dipole magnetic field 
structure of the
Earth's magnetosphere is correct since the effects of interaction 
of the Solar wind with the Earth's magnetic field are important at 
distances $\sim 8 - 10 R_{\rm Z}$. 

In order to have an idea about the importance of cascading effects which might 
be observed by different experiments under the assumption that the 
highest energy cosmic rays are photons, we estimate the probability 
of interaction of photons taking into
account the parameters of the larger number of the highest energy events.  
These probabilities are given in Table~\ref{table1} for photons with the
parameters of ten highest energy  events observed by the AGASA
array~\cite{taetal98}. The specific event in this table
is characterized by the date of  detection, its energy, the zenith angle
$Z$, the azimuth angle $A$, and the angle to the magnetic axis $\phi$. 
The
azimuth angles are measured clockwise  from the North Magnetic Pole and are
recalculated based on the  information from Table~2 in Takeda~et
al.~\cite{taetal98}.  As we can see in Table~\ref{table1}, the probability of 
interaction depends not only on the event energy but also on the
arrival direction. Therefore, the events with energies even below 
$10^{20}$
eV (e.g. the events with energy $9.8\times 10^{19}$ eV and 
$9.1\times 10^{19}$ eV), which arrive from the Northern direction, have a
higher probability  of cascading than the highest energy events detected by
the AGASA array. Moreover, the probability of interaction of  photons arriving
from the Southern direction may change significantly if the arrival directions
of photons
differ by several degrees. Therefore, the photon with the parameters of
the second highest energy event has low chance of cascading in the 
Earth's
magnetosphere (the estimated probability of cascading is only 0.04).
The cascades in the atmosphere initiated by these two highest
energy AGASA events, if photonic
in nature, should have significantly different depths of the maximum in 
the Earth's  atmosphere because of the 
influence of the  Landau-Pomeranchuk-Migdal effect~\cite{lp35,mi56,ka85}. 
Similar computations of the probability of interaction of photons 
with the parameters of ten highest energy events 
detected by the Haverah Park
array are shown in Table.~\ref{table2b}. Again, the photons arriving from the
Southern  direction may have significantly different probability of
interaction  even if their arrival direction differ by $\sim 20^{\rm o} -
30^{\rm o}$ degrees (see events with energies $1.26\times 10^{20}$ eV,
$1.16\times 10^{20}$ eV, and $9.9\times 10^{19}$ eV). These conclusions will
be  discussed further in the case of locations of the Auger Observatories
(see section~3.2).


%
%
\subsection{Cascades initiated by photons at locations of the Auger
experiment.}

The geographic latitudes of the planned Southern and Northern Auger 
Observatories are similar but their locations with respect to the 
magnetic axis 
(angles equal to $\alpha_{\rm mag}^{\rm SAO}\cong 66^{\rm o}$ for the 
Southern Observatory and to $\alpha_{\rm mag}^{\rm NAO}\cong
38^{\rm o}$ for
Northern Observatory) differ  by about $28^{\rm o}$. Therefore, it is 
expected that EHE photons arriving from specific directions to the locations of
the Southern and  Northern Auger Observatories should initiate cascades in the
magnetosphere with different characteristics. In order to have an idea 
about
the efficiency of cascading, we compute the values of the perpendicular
component of the magnetic field along different directions of motions of
primary photons, defined by the azimuth and zenith angles (see
Fig.~\ref{fig3}). In general, the dependence of the magnetic field on
direction is quite complicated. The characteristic cusps at certain directions
correspond to distances from the observatories at which the magnetic field is
oriented along the direction of primary photon. The perpendicular component of
the magnetic field reaches the lowest values for direction towards the Equator
(the azimuth angle $\phi = 180^{\rm o}$)  and
the zenith angles corresponding to the values of the angles between the
location of the specific observatory and the magnetic pole 
($Z = \alpha_{\rm mag}^{\rm SAO}$ or $Z = \alpha_{\rm mag}^{\rm NAO}$). 
The strongest perpendicular component of the magnetic field is met 
in directions
towards the magnetic poles.  Our results are not consistent with the
conclusions  presented in~\cite{sv97}. It is shown in that paper in Fig.~3
that the array on the Northern hemisphere sees the strongest
magnetic field from the Southern direction. We localize the region
of the lowest magnetic field with the Southern direction and at the 
limited values of the zenith angles (between $0^{\rm o}$ and 
$\alpha_{\rm mag}$). This result has not been found in previous 
calculations.
According to our calculations the highest energy events observed by 
the AGASA array~\cite{taetal99} come from directions where the
perpendicular component of the Earth's magnetic field is relatively low
and the cascading effects of photons should be less efficient 
in comparison to other directions.

In order to confirm our expectations, which base on the analysis of the
magnetic field profiles, we present in Fig.~6 and 7 the results of
simulations of cascades initiated by primary  photons with energies
$10^{20}$eV (figures (a) and (b)) and $3\times  10^{20}$eV (figures (c) and
(d)) arriving to the location of the Southern Auger Observatory from zenith
angles $Z = 75^{\rm o}, 60^{\rm o}, 45^{\rm o}, 30^{\rm o}, 15^{\rm o}$ and
$0^{\rm o}$ and azimuth angle $\phi = 0^{\rm o}$, measured  clockwise from the
North Magnetic Pole (full histograms from the  thickest to the thinnest in
figures (a) and (c)), and from zenith angles  $Z = 75^{\rm o}, 60^{\rm o},
45^{\rm o}, 30^{\rm o},$ and $15^{\rm o}$ and azimuth angle  $\phi = 180^{\rm
o}$ (histograms in figures (b) and (d)). Figures show the  numbers of secondary
photons per log$E_\gamma$ = 0.1 averaged over  100 simulated primary photons.
In fact, the strongest cascading effects (the higher numbers of secondary
photons), occur at the highest zenith angles in direction towards the
magnetic poles (see Figs.~\ref{fig4}a,c, and~\ref{fig5}a,c). Alternatively,
photons with energies $10^{20}$ eV arriving from the South at zenith angles
$15^{\rm o}$ and $30^{\rm o}$ do not cascade at all (see Figs.~\ref{fig4}b
and~\ref{fig5}b). Note also that photons with energies of $3\times 10^{20}$ eV
arriving to the Northern Auger Observatory at zenith angles $Z = 15^{\rm o}$
cascade more efficiently than these ones arriving at angles $Z = 30^{\rm o}$
(Fig.~\ref{fig5}d). These features are consistent with the conclusions 
obtained from the analysis of
the magnetic field profiles. It is evident that the weakest cascading
effects are for photons arriving from directions  at zenith angles somewhat
smaller than the zenith angles $Z = \alpha_{\rm mag}^{\rm SAO}$ and 
$\alpha_{\rm mag}^{\rm NAO}$. This is due to the fact that the 
perpendicular
component of the magnetic field at large distances along these directions from
the observatories is higher than for direction defined by $Z = 0^{\rm o}$
(compare the dot-dashed and dot-dot-dot-dashed curves with full curves in
Fig.~\ref{fig3}). In Fig.~\ref{fig6} we compare also the  average numbers of
secondary photons (results averaged over 100 simulated primary photons) from
cascades initiated by primary photons arriving randomly from the sky to both
Auger Observatories. Although as we show above, there are significant
differences between photons arriving from specific directions,  on average the
differences between averaged cascade spectra produced by photons arriving
randomly  from the sky to these two Observatories should be rather small. 

The secondary photons, produced in cascades in the Earth's magnetosphere, enter
the atmosphere within a very small cone. Based on deflection of the 
secondary $e^\pm$ pairs in the Earth's magnetic field during development of
the cascade, we estimate the radius of this cone on a few centimeters.
Therefore cascades initiated by separate secondary photons produced 
by this same primary photon are indistinguishable. However, 
cascades initiated by secondary photons in the atmosphere should differ 
significantly from cascades
initiated by primary photons arriving from directions with  small values of
perpendicular magnetic field because of the negligible influence of the
Landau-Pomeranchuk-Migdal (LPM) effect on the development of the photon
induced showers with energies below $\sim 10^{19}$ eV (see e.g.~\cite{ka97}). 
Therefore the showers initiated by photons arriving to the Auger 
Observatories from directions of the Equator and at the zenith angles 
between $\alpha_{\rm mag}^{\rm SAO}$ (and $\alpha_{\rm mag}^{\rm NAO}$) 
and the zenith should have a significantly larger depth of the maximum.
We plan to study these effects in details in the future paper.

\section{Photons cascading in the magnetic field of the Sun.}  

The Sun has a large scale dipole magnetic field with the  
value on the surface about an order of magnitude higher than the Earth.
Therefore it is expected that photons with energies about
an order of magnitude lower should already interact with the Solar 
magnetic field. 
The dipole magnetic moment of the Sun is $M_s\approx
6.87\times 10^{32}$ G cm$^3$. 
We neglect the influence of the active regions on the Sun's surface 
(hot spots) with 2-3 orders of magnitude stronger magnetic field, 
since they dominate only in the Solar chromosphere.

We consider the problem of whether or not secondary photons from cascades
initiated by EHE primary photons in the magnetic field of the Sun 
can be observed by detection of showers in the Earth's 
atmosphere. The motion of primary photon in the Sun's magnetosphere is
defined by the impact parameter $s$ and the angle $\phi_{\rm s}$ between
the photon path and the magnetic axis.
We investigate the cascades initiated by primary photons which enter
randomly within a circle of radius $s$ around the Sun. The number of
secondary  photons, per $\Delta log E_\gamma = 0.1$, from cascades 
initiated by primary monoenergetic  photons with  energies 
$10^{19}$ and $10^{20}$ eV, injected randomly within the circles $s = 1.5,2$ 
and $3r_{\rm s}$  around the Sun 
(where $r_{\rm s}$ is the radius of the Sun) are shown in 
Figs.~\ref{fig7}a,b by the thick, middle, and thin histograms,
respectively. These results 
are averaged over 100 simulated primary photons in the case of primary photons
with energy $10^{19}$eV, and over 10 simulations in the case of photons 
with energy $10^{20}$ eV.
All primary photons with energies $10^{19}$ eV injected within 
the circle $s = 1.5r_{\rm s}$  from the Sun cascade but only a fraction
of such photons interact if injected within larger circles 
($61\%$ for $s = 2r_{\rm s}$,  and $33\%$ for $s = 3r_{\rm s}$).
All primary photons with energies $10^{20}$ eV cascade if injected 
within the considered range of the parameter $s$. 

Let us again consider possibility that EHE cosmic ray spectrum 
$> 10^{18}$ eV contains
significant proportion of photons. We compute the spectra of 
secondary photons from cascades initiated by primary photons, with 
the cosmic ray spectrum observed at the highest energies ($\propto 
E^{-2.7}$, see e.g.~\cite{yd98}), in the magnetosphere of the Sun.
These photons are injected randomly within a certain circle $s$ around
the Sun. In Fig.~\ref{fig8}a, we show the spectra of secondary 
photons (multiplied by the photon energy
squared) from cascades initiated by primary EHE photons injected within 
$s = 1.5, 2, 3 r_{\rm s}$ (from the thickest to thinnest histograms
respectively). It is assumed that the primary photon spectrum extends 
up to $E_{\rm max} = 3\times 10^{20}$ eV. It is normalized to the 
observed cosmic ray spectrum at $10^{19}$ eV. The observed cosmic ray
spectrum at energies above $10^{14}$eV and the spectrum of injected
primary photons are marked in this figure  by the dashed and dotted curves, 
respectively. As expected, the spectrum of primary photons cut-offs at 
lower
energies for smaller impact parameters (defined by $s$), as a result of
cascading in the Sun's magnetic field. For the circle around the Sun defined 
by $s = 1.5r_s$, photons with  energies above $\sim 10^{19}$eV should  not be
observed,  but for $s = 3r_s$ only photons with energies above  $\sim
10^{20}$eV are absorbed. In Fig.~\ref{fig8}b we show that the spectrum of
secondary photons is almost independent on the cut-offs in the  primary
spectrum. However such dependence may be present if the
spectrum of primary photons above a few $10^{20}$eV flattens 
considerably as predicted by
some exotic theories of cosmic ray origin~\cite{bs98}.   

The solid angle corresponding to a circle with the radius 
$2r_{\rm s}$ around the Sun is equal to
$\Delta\Omega = 3r_{\rm s}^2/4r_{\rm s-z}\approx 1.6\times  10^{-5}$ sr
(Sun disk discounted), 
where $r_{\rm s-z}$ is the distance of the Earth from the Sun.
Therefore the chance of detection of a particle with energy $> 10^{19}$eV from
the direction of the Sun by the present (and past) detectors is very low.
The number of events which will be detected by both 
Auger Observatories can not be precisely predicted, since the 
spectrum of cosmic rays at the highest energies is not well known. 
We assume, following Boratav~\cite{boetal97}, that both
Auger Observatories will observe several thousands events
$> 10^{19}$ eV during one year of  operation. The number of such
events from the circle of 2$r_s$ around the Sun observed by both
Auger Observatories can be estimated from 
\begin{eqnarray}
N = k \Delta\Omega N_{\rm AO}\approx 1.2~{\rm events~per~10~years},
\end{eqnarray}  

\noindent
where $k\approx 0.5$ is the part of the sky observed by the 
Auger Observatories, and $N_{\rm AO}\approx 15000$ is the
number of events which are likely to be observed by the Auger 
Observatories above $10^{19}$ eV. 
Therefore, we should expect some events from the direction of the Sun
during years of operation of the Auger Observatories.
If photons dominate the cosmic ray spectrum above $\sim 10^{19}$ eV,
the secondary photons from their cascades in the Sun's magnetosphere
arrive to the Earth's atmosphere with the significant perpendicular 
extent which is the result of deflection of paths of the secondary 
electrons by the Sun's magnetic field. Using our cascade code we 
have estimated this perpendicular extent by  counting
deflection of secondary cascade electrons from 
the direction of the primary photon. The
angles of deflection of escaping secondary photons are estimated from
\begin{eqnarray}
\theta\approx \sum_{\rm i}\Delta x_{\rm i}/r_{\rm L},
\end{eqnarray}
\noindent
where the sum is over all secondary parent electrons (and positrons)
which are responsible for production of specific escaping secondary 
photon, $\Delta x_{\rm i}$ is the propagation distance of i-th secondary
electron (or positron) responsible for production of secondary photon 
of i-th generation, and $r_{\rm L}$ is the Larmor radius of secondary 
electron (or positron) in the local magnetic field of the Sun's 
magnetosphere. The fact that secondary electrons and positrons of 
the cascade are deflected
in opposite directions is included in the calculations by taking the
deflection of positrons with opposite sign than in the case of 
electrons. In Fig.~\ref{fig9} we show the number of secondary 
photons with energies above $E_{\gamma, s}$, which fall on the 
Earth's atmosphere within a ring with the width $\Delta D = 0.2$ km and
at the distance $D$ from direction of the primary photon. 
Figure (a) shows the  results for the  primary 
photon with energy $E_\gamma = 10^{20}$ eV, entering the Sun's
magnetosphere at the distance $s = 2r_{\rm s}$,  and at the 
angle $\phi_{\rm s} = 0^{\rm o}$ and for $E_{\gamma ,s} = 10^{18}$ eV,
$10^{17}$ eV,  and $10^{16}$ eV (from the thickest to the thinnest histogram).
Figure  (b) shows the results for $E_\gamma = 10^{20}$ eV and 
$\phi_{\rm s} = 0^{\rm o}$, and $E_{\gamma ,s} = 10^{17}$ eV, but for 
$s = 4, 3, 1.1R_{\rm s}$  (from the thickest to the thinnest histogram).
Figure (c) shows the results for $E_\gamma = 10^{19}$ eV, $E_{\gamma ,s} =
10^{17}$ eV and:   $s = 1.5r_{\rm s}$ and $\phi_{\rm s} = 0^{\rm o}$ 
(thick histogram),  $s = 1.1R_{\rm s}$  and $\phi_{\rm s} = 90^{\rm o}$
(middle), $s = 1.5R_{\rm s}$  and $\phi_{\rm s} = 90^{\rm o}$ (thin). 
As
expected, the perpendicular extent  of secondary photons depends strongly on
their energies. Secondary photons with energies above $10^{17}$ eV
create 'a
core' (the number of secondary photons per ring greater than 3 for 
primary photons with energy $E_\gamma = 10^{20}$ eV  and greater than 
2 for $E_\gamma
= 10^{19}$ eV) with a typical extent of a few kilometers, and an 
extended 'tail' with secondary photons sporadically reaching distances 
even a few tens kilometers from direction of the primary photon. 
Note that the perpendicular extent of secondary photons will mainly 
concern
one direction which is perpendicular to the direction of the magnetic 
field in the place of  the cascade in the Sun's magnetosphere. 
This is due to the ordered magnetic field of the Sun. Therefore
it is expected that the secondary photons fall 
on the Earth's atmosphere within a highly prolate 
ellipse. This feature may help to distinguish such events from 
the events which produce ordinary circular showers.

The cascade code allows us also to estimate the time delay between 
the arrival of specific secondary photons to the Earth's atmosphere. 
It results from the fact that the velocities of secondary electrons 
responsible for
emission of specific secondary photons are different and because 
of different path lengths of secondary photons from the Sun to the 
Earth's atmosphere. The the time delay due to the development of the 
cascade can be estimated from
\begin{eqnarray}
\Delta t_{\rm k} = \sum_{\rm i} \Delta x_{\rm i} (1 - \beta)/c\beta
\approx \sum_{\rm i}\Delta x_{\rm i} /2c\beta \gamma^2, 
\end{eqnarray}
\noindent
where $\beta$ and $\gamma$ are the velocity and the Lorentz factor of
secondary electrons, and $c$ is the velocity of light. Since the 
Lorentz factors of electrons which produce secondary photons with 
investigated energies are very high (above $\sim 10^{10}$), this time 
delay is negligible. The time delay  due to
the propagation  of secondary photons from the Sun  to the Earth can be
estimated from 
\begin{eqnarray}
\Delta t_{\rm p} = r_{\rm s-z}(1/\cos\theta - 1)/c\approx r_{\rm s-z}
\theta^2/2\approx D^2/2cr_{\rm s-z}\approx 1.1\times 10^{-14} D^2 
{\rm ~s}, 
\end{eqnarray}
\noindent
where $D$ 
is in kilometers. Therefore for photons falling at large distances this 
time delay might be measurable.

The Auger Observatories will be arrays of 10 m$^2$ Water Cherenkov 
detector units in a 1.5km spaced triangular grid. 
The synchronous, multiple secondary photons  
with energies above $10^{17}$ eV,
produced by primary photons with energies above $10^{19}$ eV in the Sun's
magnetosphere, initiate separate showers in the atmosphere within
the radius of $\sim 2$ km. In the case of primary photons
with energies of $10^{20}$ eV the core radius have the perpendicular 
extent of
a  few kilometers (see Fig.~\ref{fig9}). Each
shower should trigger at least one detector with the signal of $\sim 3$
vertical equivalent muons (VEM) (see Fig.~4.8 in~\cite{auger}), and some nearby
detectors with the signal below 1 VEM. As we noted, the triggered
detectors should be distributed within prolate ellipse which should 
help to distinguish such showers from the ordinary circular showers.
We 
suggest that it is worth to investigate the future Auger Observatories 
data for such extraordinary showers from the direction of the Sun 
because they may give interesting information about the photon content 
in the highest energy cosmic rays.

\section{Summary and Conclusion.} 

We have investigated the interaction of EHE photons with the magnetic 
field of the Earth and the Sun in the context of detection 
capabilities of 
products of such magnetic $e^\pm$ pair cascades by the past, present, 
and future  detectors of the EHE cosmic ray showers. It is assumed that
photons may be copious at the highest energies as suggested by some
models of EHE cosmic ray origin.

Assuming that the highest energy showers observed by different cosmic
ray detectors (AGASA, Fly's Eye, Yakutsk, and Haverah Park) are caused 
by photons, we estimate their probability of cascading in the Earth's
magnetic field and compute the spectra of secondary photons which
fall on the Earth's atmosphere. We find that the probability of 
cascading
strongly depends not only on the photon energy but also on the arrival
direction of the  photon (see Tables~\ref{table1},\ref{table2b}).
Therefore the highest energy  events detected by the AGASA and Haverah
Park arrays do not have the highest probability of cascading when compared
with other events detected by these arrays. Our computations show that 
the
fluctuations in the number of secondary cascade photons are high. 
In the case
of photons with the parameters of the highest energy events detected 
by the
AGASA array, there is a small probability that cascades may
not even develop  (see Figs.~\ref{fig2}a,d, and
Tables.~\ref{table1} and~\ref{table2b}). However photons with
parameters of the highest energy events observed by the  Haverah Park,
Yakutsk, and Fly's Eye detectors should cascade with the probability 
higher than $99\%$. These effects are caused by the fact that the 
perpendicular component of the magnetic field along the path of the
primary photon strongly depends on the direction and location of the
detector (Fig.~5). We have found that for detector located on the 
Northern hemisphere the highest probability of cascading is for photons
arriving from the Northern
direction and the lowest for the Southern direction at the zenith angles
between the zenith and  $\alpha_{\rm mag}$, where $\alpha_{\rm mag}$ is the
angle between the location of the detector and the magnetic axis. 
This region
of the weakest magnetic field and so the lowest probability of cascading has
not been localized in the previous works on this topic. The 
probability of initiating a cascade close to the border of this 
region can change drastically even if the direction of motion of the 
primary photon changes by several degrees. Therefore two photons with
comparable energies but arriving within the angular
distance of several degrees from the direction of the equator to the 
detector may behave differently (e.g. AGASA events with energies 
$2.1\times 10^{20}$ and $1.5\times 10^{20}$ eV or Haverah Park events
with energies $9.9\times 10^{19}$ eV and $1.16\times 10^{20}$ eV).

The above conclusions are
verified by simulation of cascades initiated by photons arriving to
the locations of the planned Southern and Northern Auger Observatories 
(see Figs.~\ref{fig4},\ref{fig5}). As predicted the least efficient 
cascading of photons occurs for directions from the Equator and at
specific range of zenith angles. Since the angles between the location of 
these Observatories to the magnetic axis will differ by about 
$28^{\rm o}$ degrees, the regions of the lowest probability of 
cascading are different. However the spectra of
secondary photons averaged over random directions of 
primary photons with energies $10^{20}$ and
$3\times 10^{20}$ eV do not show large
differences between the Southern and Northern Auger Observatories
(Fig.~\ref{fig6}). Therefore, the possibility of cascading by photons 
with parameters of specific EHE events has to be considered individually.

The secondary photons produced in cascades arrive at 
the Earth's atmosphere within a very small cone with
the radius of the order of centimeters. Therefore cascades initiated in the 
atmosphere by specific secondary photons are indistinguishable. However, the
energies of secondary photons are significantly lower than those  
of primary photons. This has consequences for the development of cascades 
in the atmosphere because of the lower influence of the
Landau-Pomeranchuk-Migdal effect. Therefore the showers with this same energy
but arriving from regions on the sky which we call the 'high' and 'low'
probability regions, should have on average different depths of the
maximum. If photons are common at the highest energies, then such differences
of the depth of the maximum of the highest energy cosmic ray showers should be
observed by the Auger Observatories. 

In the future paper we plan to
investigate the cascades initiated in the Earth's atmosphere by photons
with the parameters of the highest energy events detected by
the operating detectors using as an input the results presented in this 
work. It is likely that energies of these events, estimated based on 
the development of hadronic cascades in the atmosphere, may not give
good estimate for the energy of primary particle under the hypothesis 
that photon is responsible for the specific event. We intend to obtain
the reported detector response to the observed highest energy showers
by searching for the best energy of the primary photon. 
Such simulations in the Earth's atmosphere will be done also for 
photons arriving from different directions to both Auger Observatories.

We have investigated the consequences of cascades
initiated by EHE photons with energies above $\sim 10^{19}$ eV
in the magnetosphere of the Sun. The secondary photons, products of
such cascades, arrive to the Earth's atmosphere within a highly
prolate ellipse with the 'core' dimension of the main axis of the order
of a few kilometers (see Fig.~11). The 'core' dimension strongly depends
on the energy of secondary photons, i.e. the higher energy photons
arrive closer to the direction of the primary photon. 
Some secondary photons from
such cascades can even trigger detectors of the array at the distance
of a few tens kilometers. These secondary photons arrive to the Earth
almost simultaneously (see Eqs.~(3) and (4)). We predict that if photons 
are common at the highest energies, then both Auger
Observatories may observe some very atypical showers from the direction 
of the Sun during years of operation.  
The future Auger Observatory data should be investigated
for very extraordinary
showers coming from the direction of the Sun.
The predicted rate of such showers will be low (about 1 per 10 years). 
However it will be possible to distinguish them from the ordinary 
circular showers due to the different shape and lateral profile.
In the future work we plan to perform simulations of the cascades 
initiated in the Earth's atmosphere by secondary photons produced
in the cascade initiated by primary photon in the Sun's magnetic field.

\section*{Acknowledgments.}
\noindent
I would like to thank the anonymous referees for useful comments and 
suggestions, and  Prof. M. Giller and dr J. H. Beall for discussion
and reading the manuscript. 
This work is supported by the {\it Komitet Bada\'n Naukowych} grant  
No. 2P03D 001 14.

\newpage

\begin{table}
\begin{center}
\caption{Probability of cascading of photons with parameters of the 
AGASA events.   
\label{table1}}
\begin{tabular}{llllll}
\hline \hline
shower  & $E_\gamma$ & Z &  $A$ & $\phi$
& Prob.  \\ \hline 
94/07/06 & 1.06 & $40.4^{\rm o}$ &  $37^{\rm o}$ & $34.2^{\rm o}$ &1.0 \\ 
84/12/17 & 0.98 & $30.4^{\rm o}$ &  $52^{\rm o}$ & $45.4^{\rm o}$ &1.0 \\ 
91/11/29 & 0.91 & $45.0^{\rm o}$ &  $14^{\rm o}$ & $19.5^{\rm o}$ &1.0 \\ 
96/10/22 & 1.05 & $37.7^{\rm o}$ &  $101^{\rm o}$& $72.5^{\rm o}$ &0.99 \\ 
93/12/03 & 2.13 & $22.5^{\rm o}$ &  $205^{\rm o}$& $80.1^{\rm o}$ &0.97 \\ 
92/09/13 & 0.93 & $25.6^{\rm o}$ &  $70^{\rm o}$ & $54.4^{\rm o}$ &0.97 \\ 
96/01/11 & 1.44 & $30.2^{\rm o}$ &  $122^{\rm o}$& $77.4^{\rm o}$ &0.96 \\ 
98/06/12 & 1.20 & $27.5^{\rm o}$ &  $140^{\rm o}$& $81.5^{\rm o}$ &0.54 \\ 
93/01/12 & 1.01 & $33.4^{\rm o}$ &  $224^{\rm o}$& $85.1^{\rm o}$ &0.4 \\ 
97/03/30 & 1.50 & $43.6^{\rm o}$ &  $188^{\rm o}$& $102.5^{\rm o}$&0.04 \\

\hline \hline
\end{tabular}
\end{center}
\end{table}

\newpage

\begin{table}
\begin{center}
\caption{Probability of cascading of photons with parameters of the 
Haverah Park events.   
\label{table2b}}
\begin{tabular}{llllll}
\hline \hline
shower  & $E_\gamma$ & Z &  $A$ & $\phi$ &
Prob.  \\ \hline 
5746080 & 1.58  & $57^{\rm o}$ &  $5^{\rm o}$  & $16.4^{\rm o}$ &1.0 \\
5140783 & 1.15  & $26^{\rm o}$ &  $297^{\rm o}$& $36.1^{\rm o}$ &1.0 \\ 
17684312 & 1.01 & $37^{\rm o}$ &  $50^{\rm o}$ & $31.1^{\rm o}$ &1.0 \\ 
9502797  & 0.81  & $41^{\rm o}$ &  $358^{\rm o}$&$1.7^{\rm o}$  & 1.0 \\ 
10643668 & 0.75  & $49^{\rm o}$ &  $17^{\rm o}$ &$14.4^{\rm o}$&1.0 \\ 
9160073  & 1.59  & $30^{\rm o}$ &  $109^{\rm o}$&$56.3^{\rm o}$&0.99 \\ 
10195820 & 0.99  & $51^{\rm o}$ &  $140^{\rm o}$&$85.2^{\rm o}$&0.89 \\ 
12701723 & 1.26 & $29^{\rm o}$ &  $226^{\rm o}$ &$64.0^{\rm o}$&0.49 \\ 
9597348  & 0.72  & $11^{\rm o}$ &  $82^{\rm o}$ &$40.8^{\rm o}$&0.27 \\ 
8185176  & 1.16  & $35^{\rm o}$ &  $210^{\rm o}$&$73.1^{\rm o}$&0.26 \\ 
14617294 & 0.63  & $42^{\rm o}$ &  $182^{\rm o}$&$83.1^{\rm o}$&0.0 \\ 
12953265 & 0.55  & $14^{\rm o}$ &  $112^{\rm o}$&$47.8^{\rm o}$&0.0\\ 
\hline \hline
\end{tabular}
\end{center}
\end{table}



%
%

\begin{figure} 
\vspace{19.truecm} 
\includegraphics{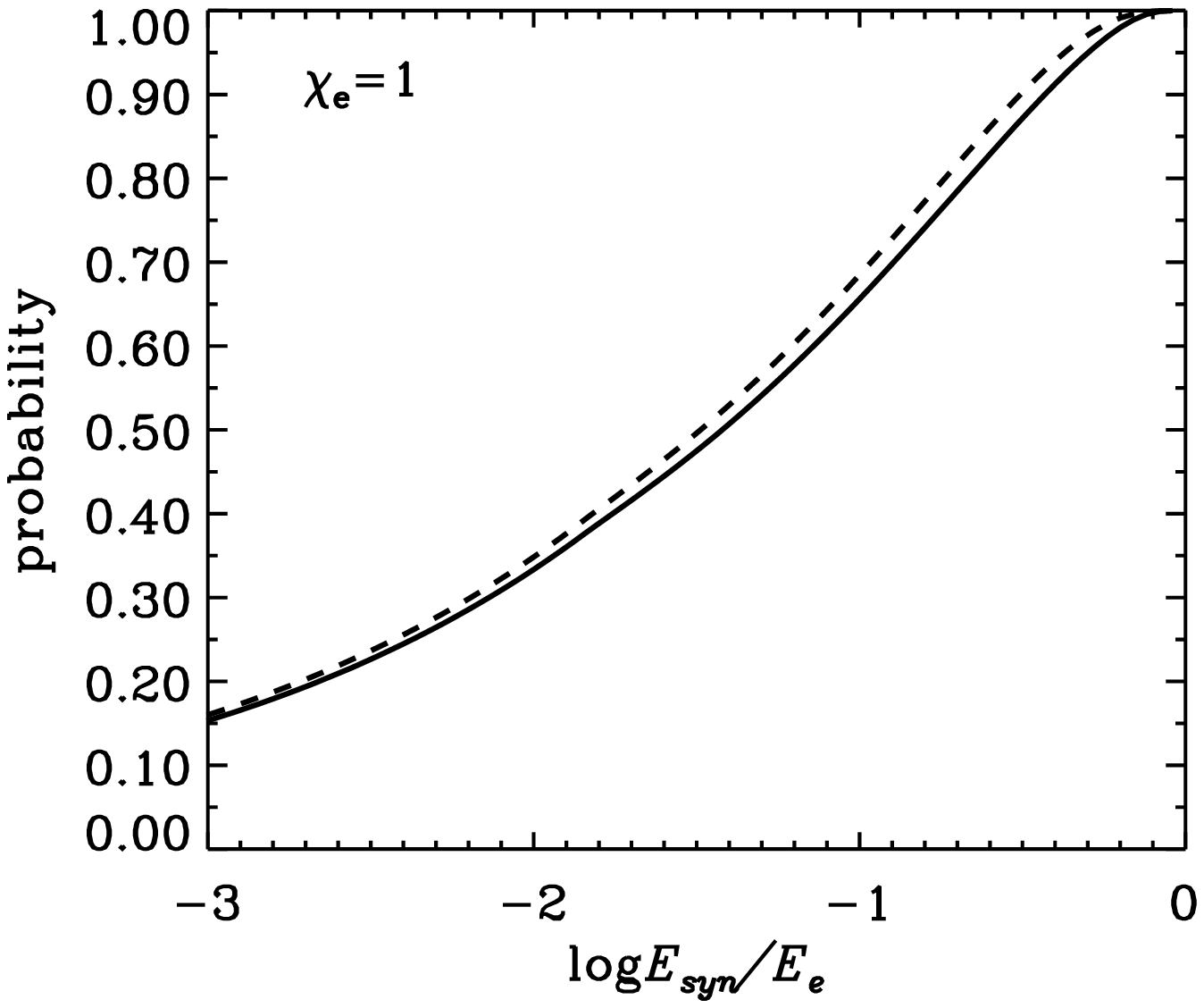}
\includegraphics{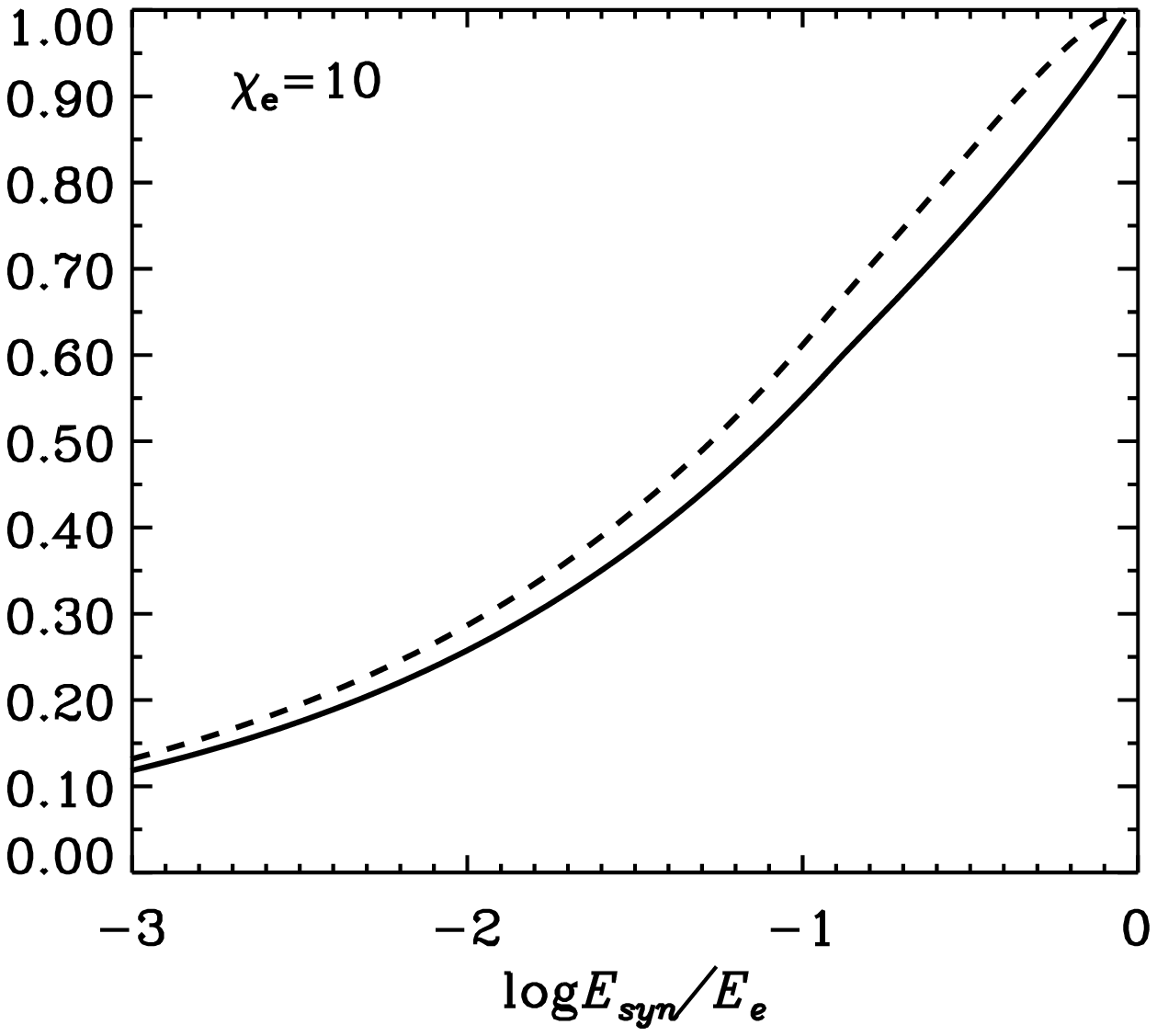}
\includegraphics{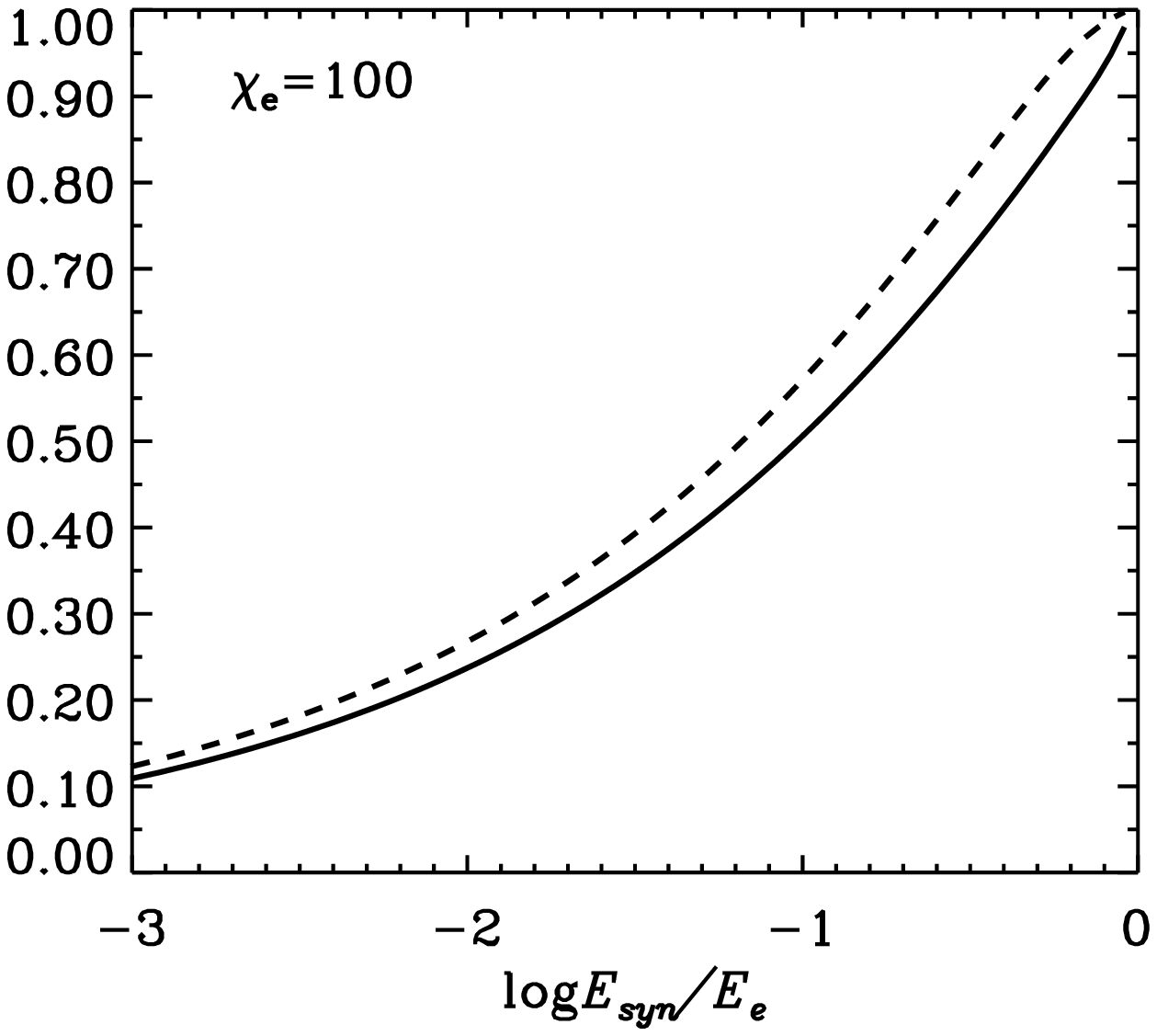}
\includegraphics{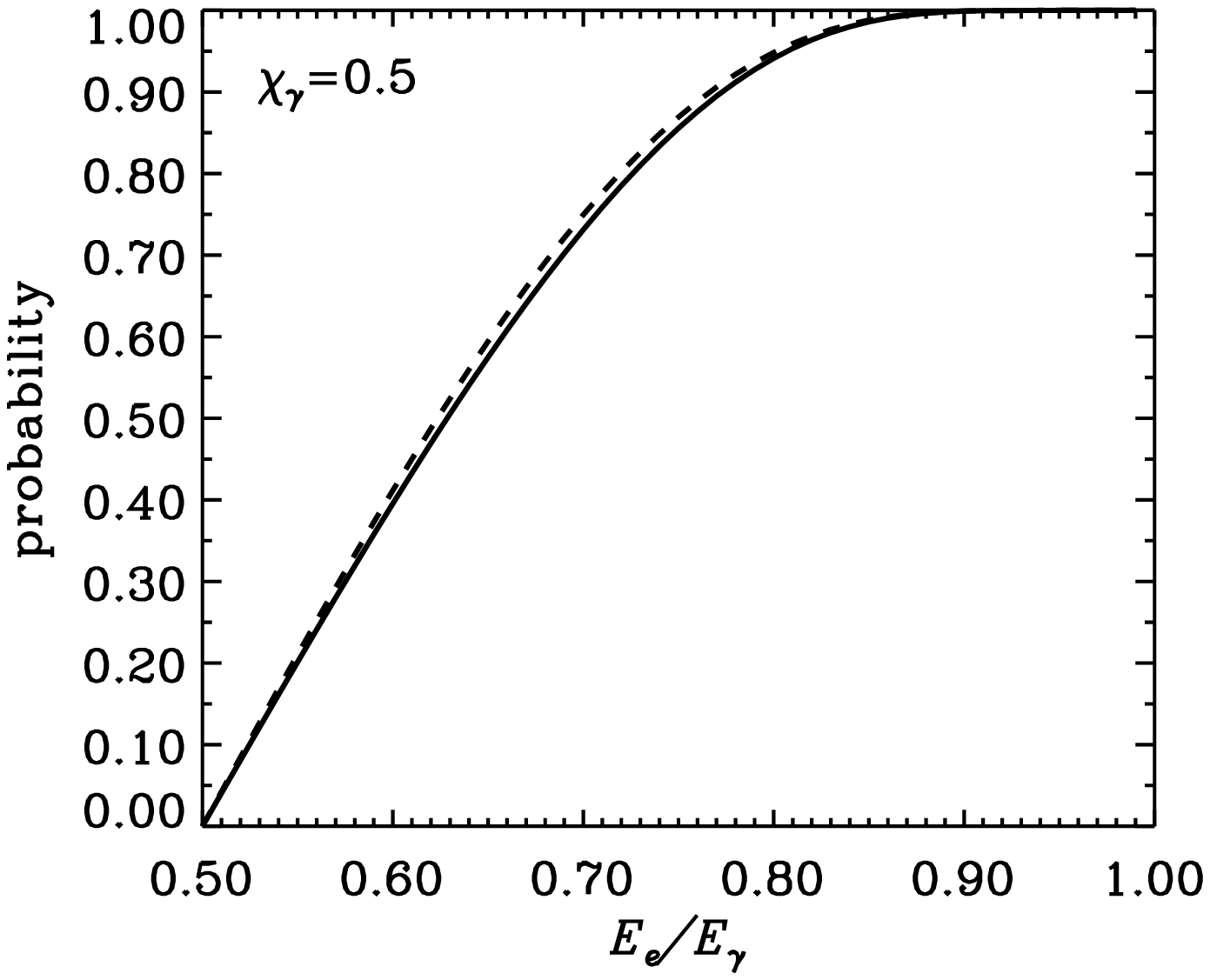}
\includegraphics{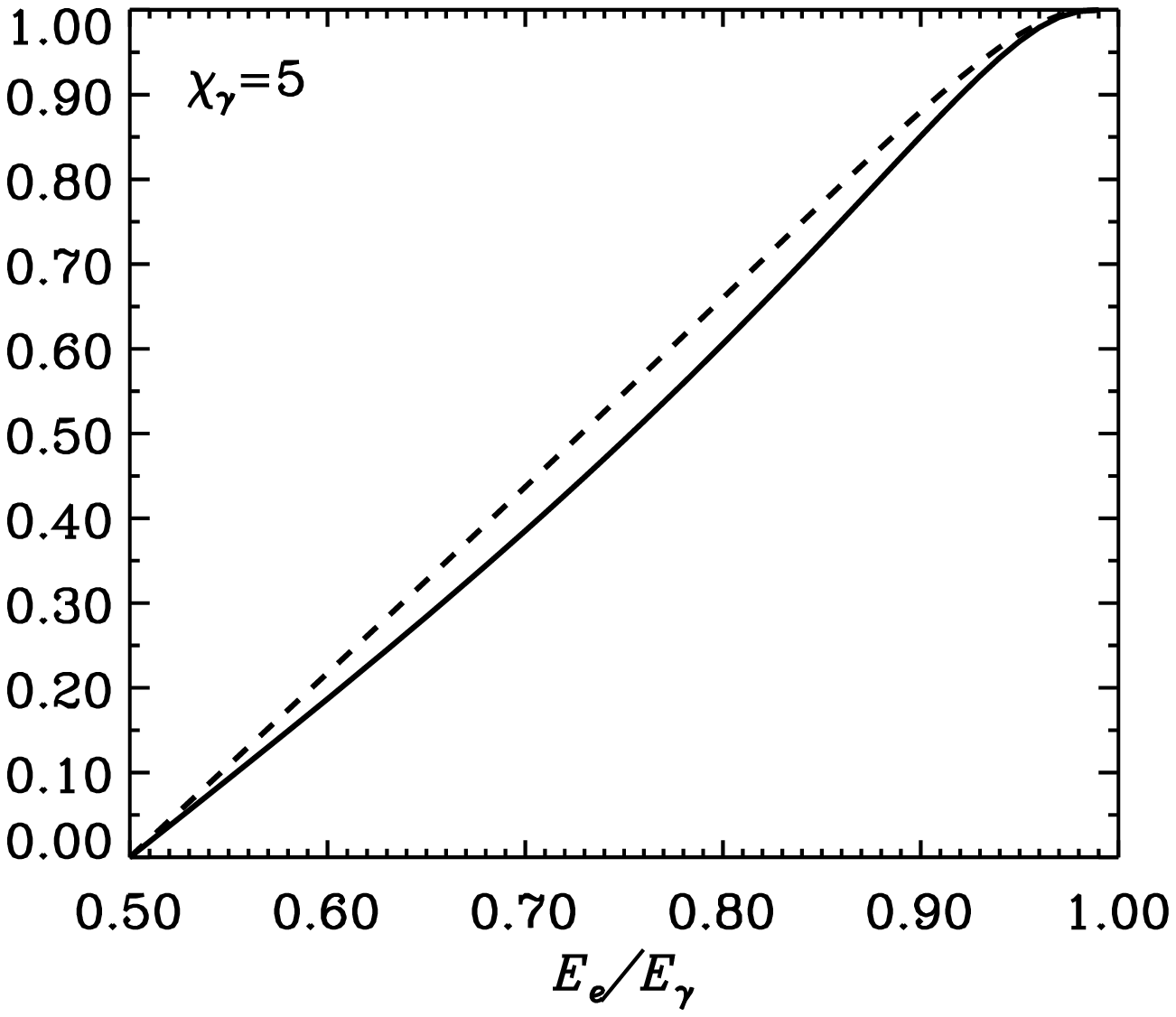}
\includegraphics{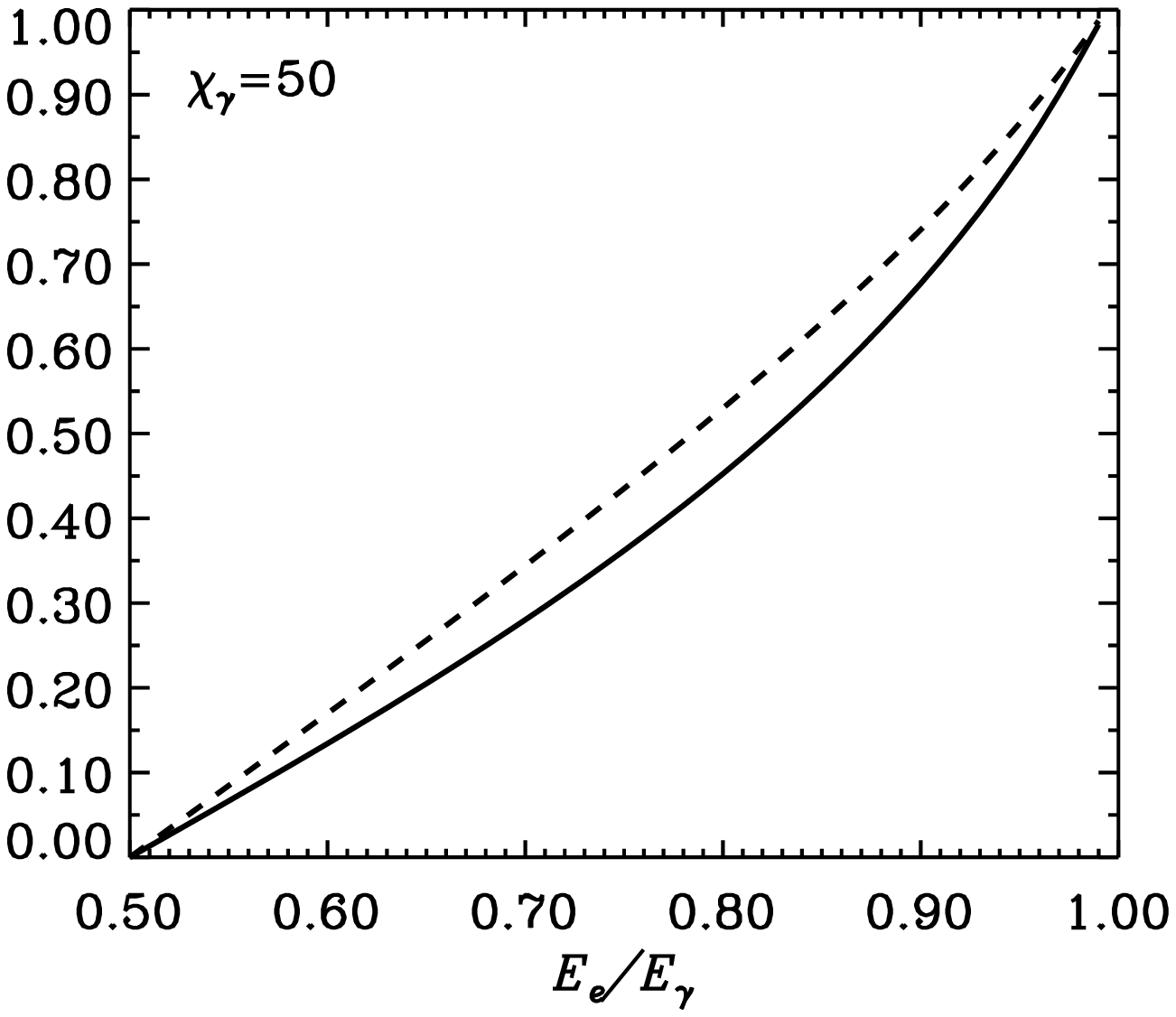}         
\caption[]{Probabilities of production of an $e^\pm$ pair
with energy below $E_{\rm e}$ by  the photon with energy $E_\gamma$
(top figures) and the probabilities of emission of synchrotorn photons
$E_{\rm syn}$ by the electron with energy below $E_{\rm e}$ (bottom 
figures) 
in the magnetic field corresponding to the selected values of the 
parameter $\chi_\gamma$ and  $\chi_{\rm e}$. The computations with the
Baring's rates are shown by the full curves and with the Erber's 
rates by the dashed curves.}   
\label{fig0} 
\end{figure}  

\begin{figure} 
\vspace{6.truecm} 
\includegraphics{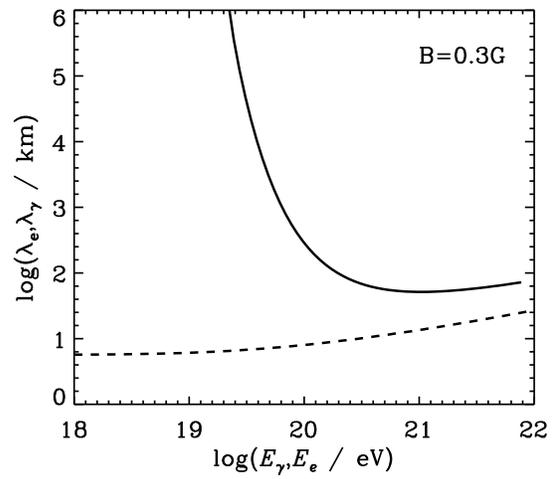}   
\caption[]{
The mean free path for conversion of photon with
energy $E_\gamma$ into $e^\pm$ pair (full curve) and for production of
synchrotron photon by electron (positron) with energy $E_{e}$
in the perpendicular magnetic field with strength 0.3 G.} 
\label{fig1} 
\end{figure} 

\begin{figure} 
\vspace{19.truecm} 
\includegraphics{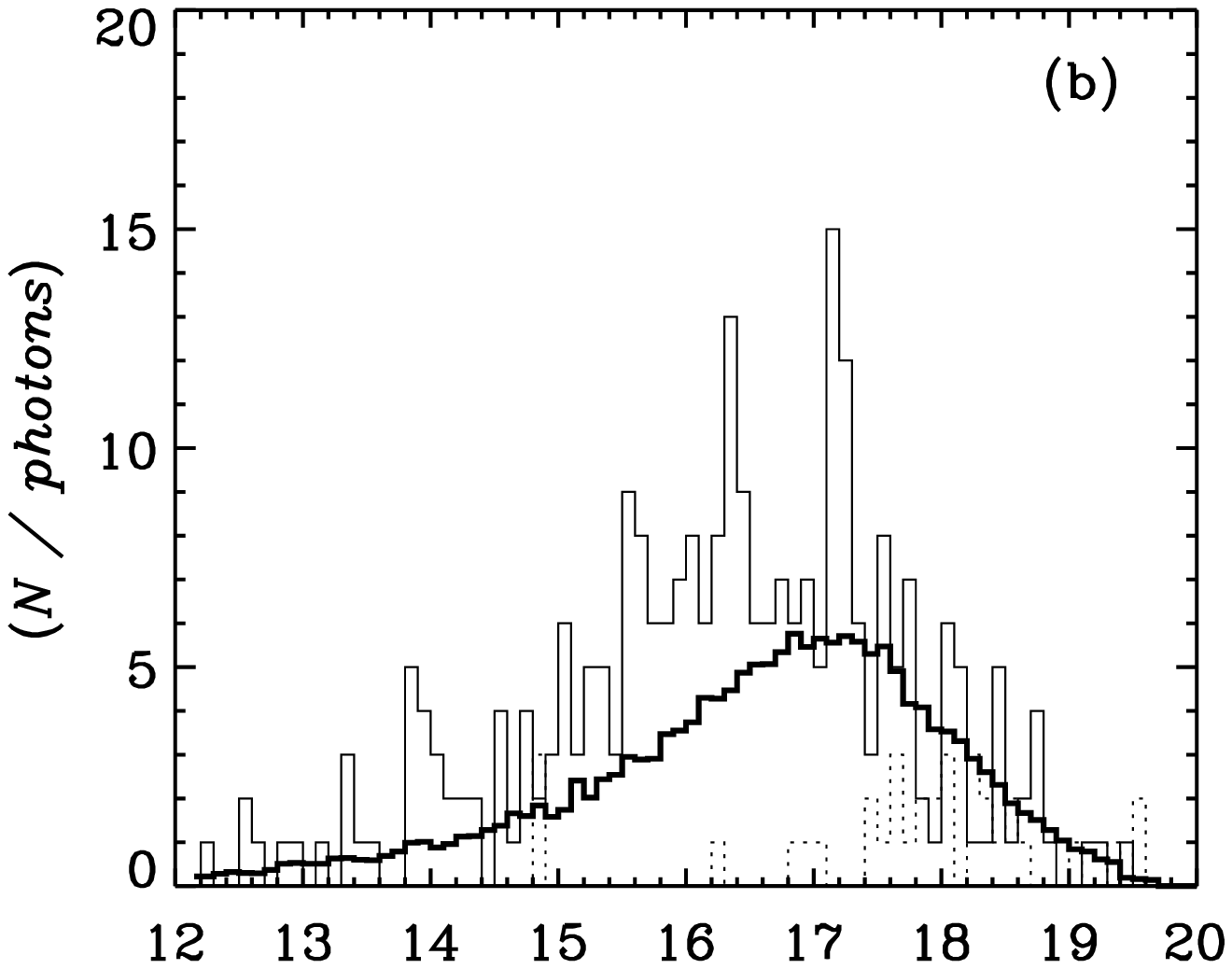}
\includegraphics{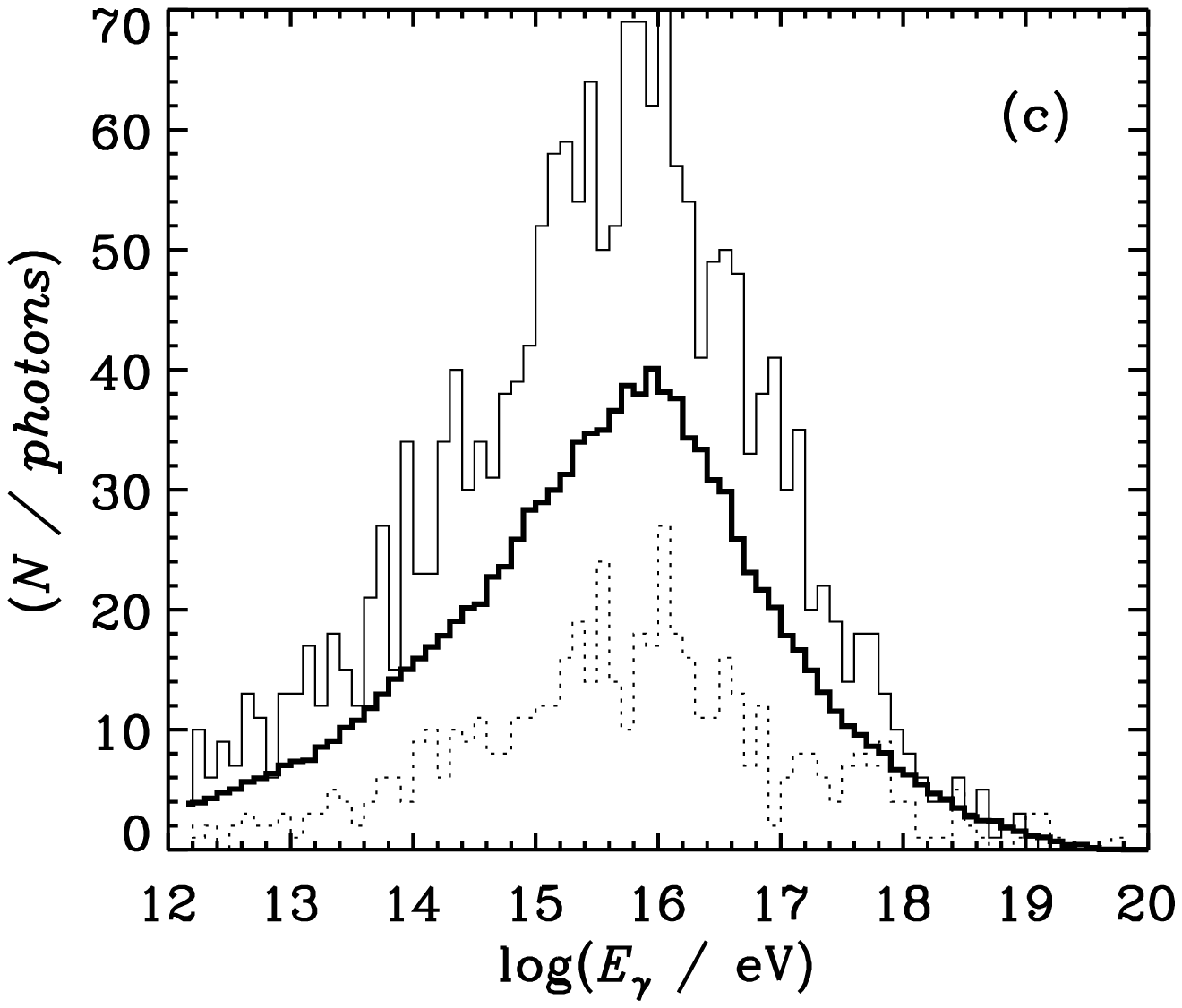}
\includegraphics{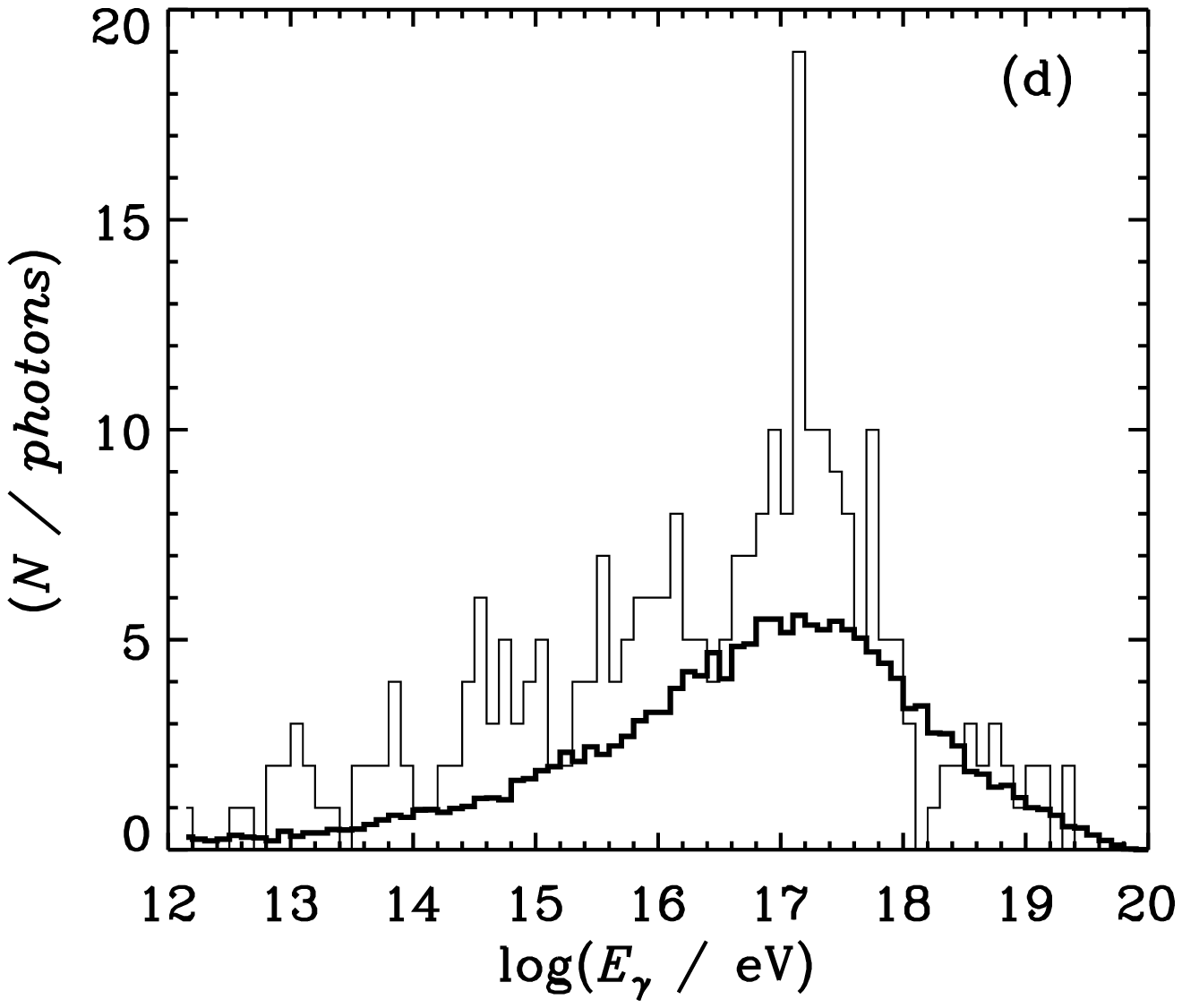}
\includegraphics{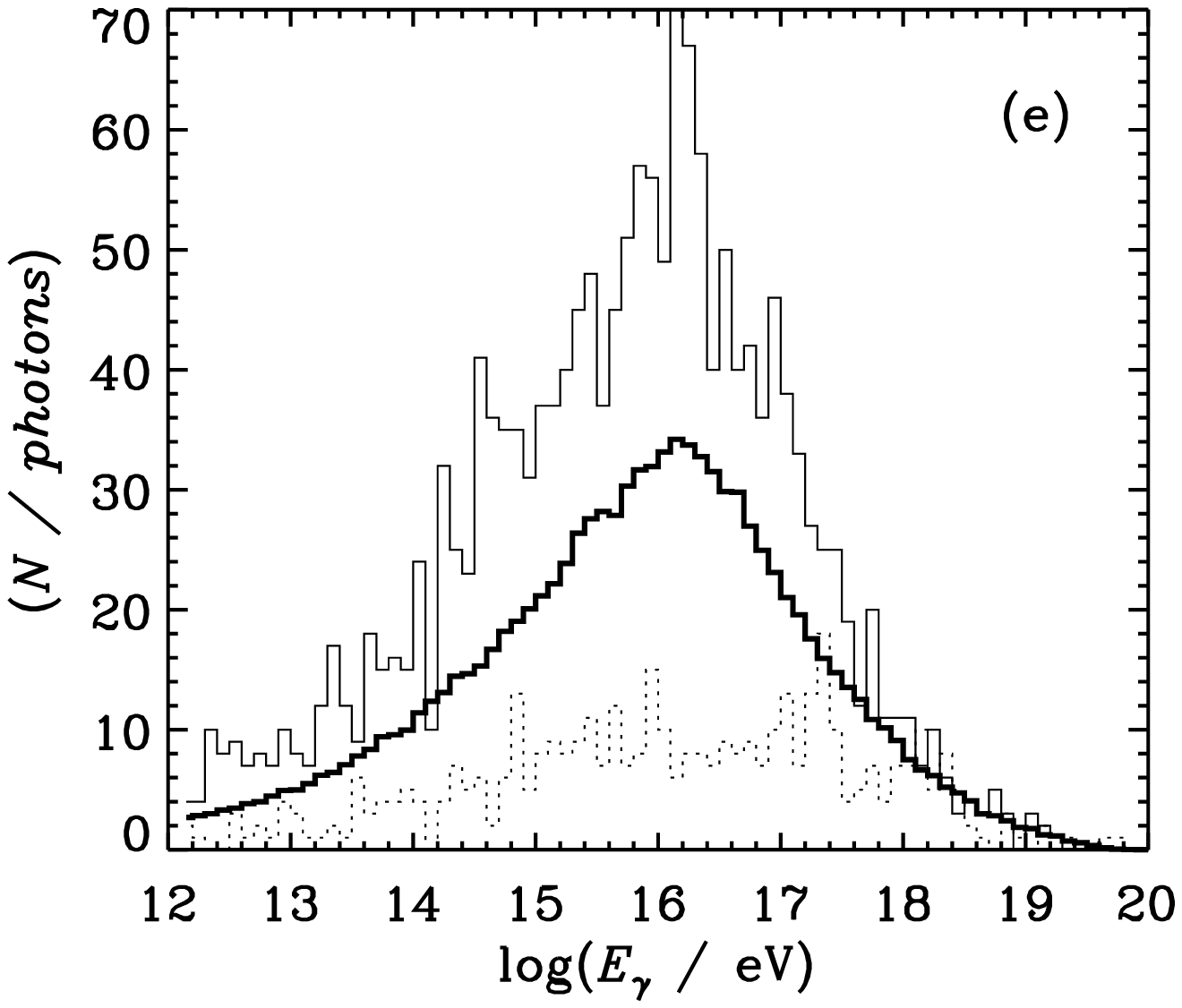}         
\caption[]{
Numbers of secondary photons, per $\Delta(log E_\gamma) = 
0.1$, produced in cascades in the Earth's magnetic field which are
initiated by primary photons with the parameters (energies
and arrival directions) estimated for the highest energy events 
observed by: (a) the Haverah Park array (shower 9160073 with energy of
$1.59\times 10^{20}$ eV); (b)  the Yakutsk array 
(event with energy of $2.3\times 10^{20}$ eV); (c) the AGASA array (shower
93/12/03 with energy of $2.1\times 10^{20}$ eV); (d) the Fly's Eye
Observatory (shower with energy of $3.2\times 10^{20}$ eV) . The full
histograms show results averaged over 100 simulated primary photons. The
cascades with the maximum and minimum number of secondary photons in our 100
simulations are shown by the thin full and dotted histograms, respectively.}   
\label{fig2} 
\end{figure}  

\begin{figure} 
\vspace{10.truecm} 
\includegraphics{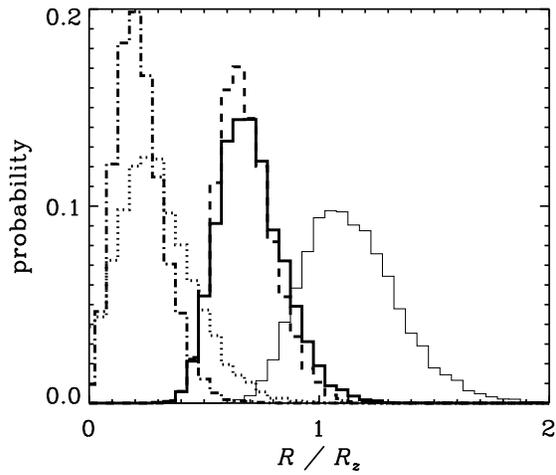}   
\caption[]{Probability ($\Delta N/\Delta log(R/R_{\rm Z}))$ of 
the first interaction of photons discussed in Fig.~\ref{fig2} as a 
function of distance from  the surface of the Earth measured in  
units of the Earth's
radius. Specific histograms show the probability for events detected by:
Haverah Park (dot-dashed), Yakutsk (dashed), AGASA (dotted), Fly's Eye 
(thick full). The thin full histogram  shows the probability for the 
case of supposed event with the arrival parameters of the Fly's Eye 
event but with the energy $10^{21}$ eV. All results are
averaged over $10^4$ simulations.}  
\label{fig11}  
\end{figure} 

\begin{figure} 
\vspace{10.truecm} 
\includegraphics{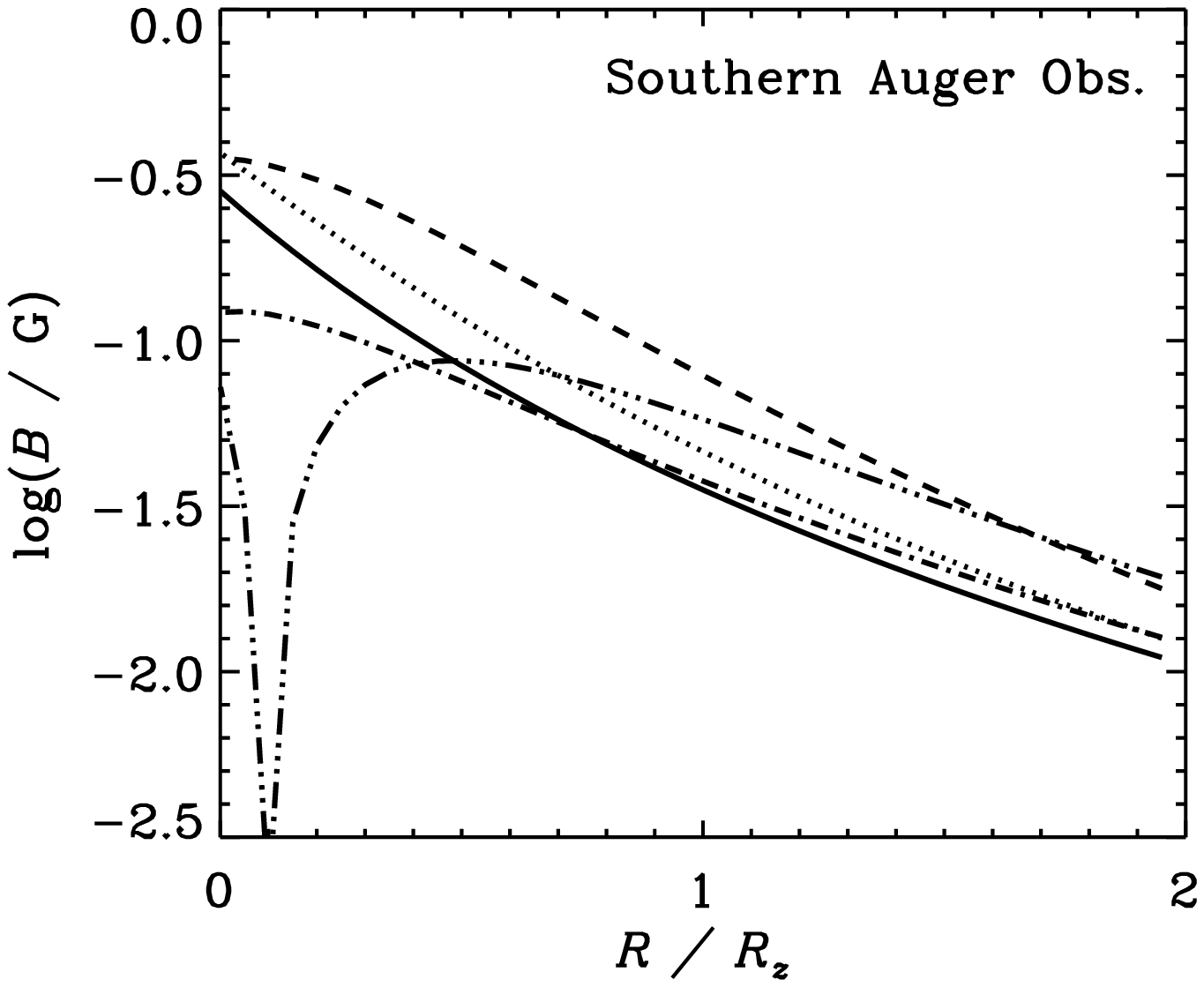}
\includegraphics{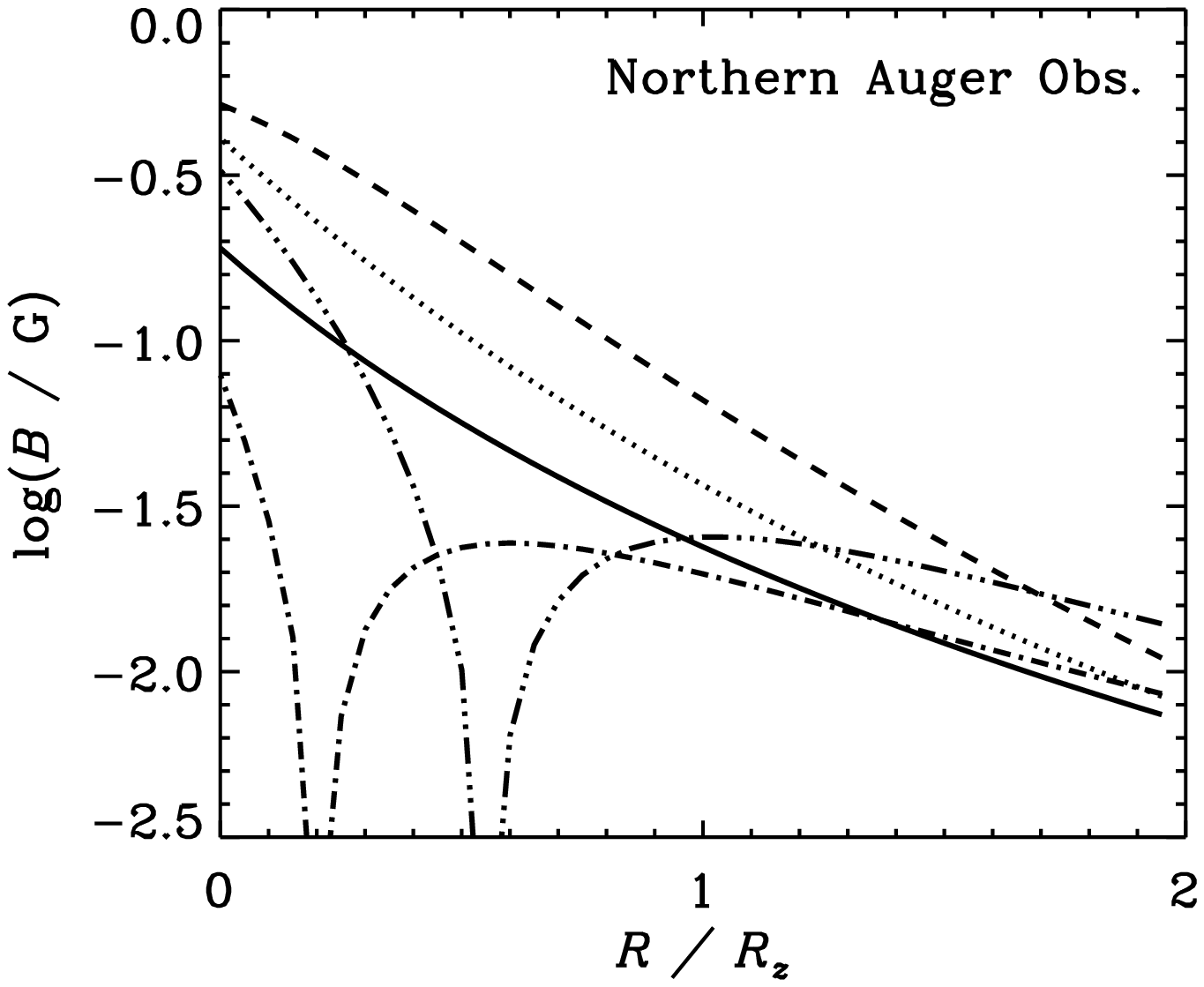}      
\caption[]{
Magnetic field profiles along direction of motion of EHE photons,
defined by the azimuth angle $\phi$ and the zenith angle $Z$, 
at locations of the Southern and Northern Auger Observatories. 
The distance from the observatories is given in units of the radius of the
Earth $R_z$. Separate curves correspond to the following directions:
$Z = 0^{\rm o}$ (full); $\phi = 0^{\rm o}$, and $Z = 30^{\rm o}$ (dotted),    
and $Z = 60^{\rm o}$ (dashed); and $\phi = 180^{\rm o}$, and $Z = 30^{\rm o}$
(dot-dashed), and $Z = 60^{\rm o}$ (dot-dot-dot-dashed).}   
\label{fig3}
\end{figure} 

\begin{figure} 
\vspace{12.truecm} 
\includegraphics{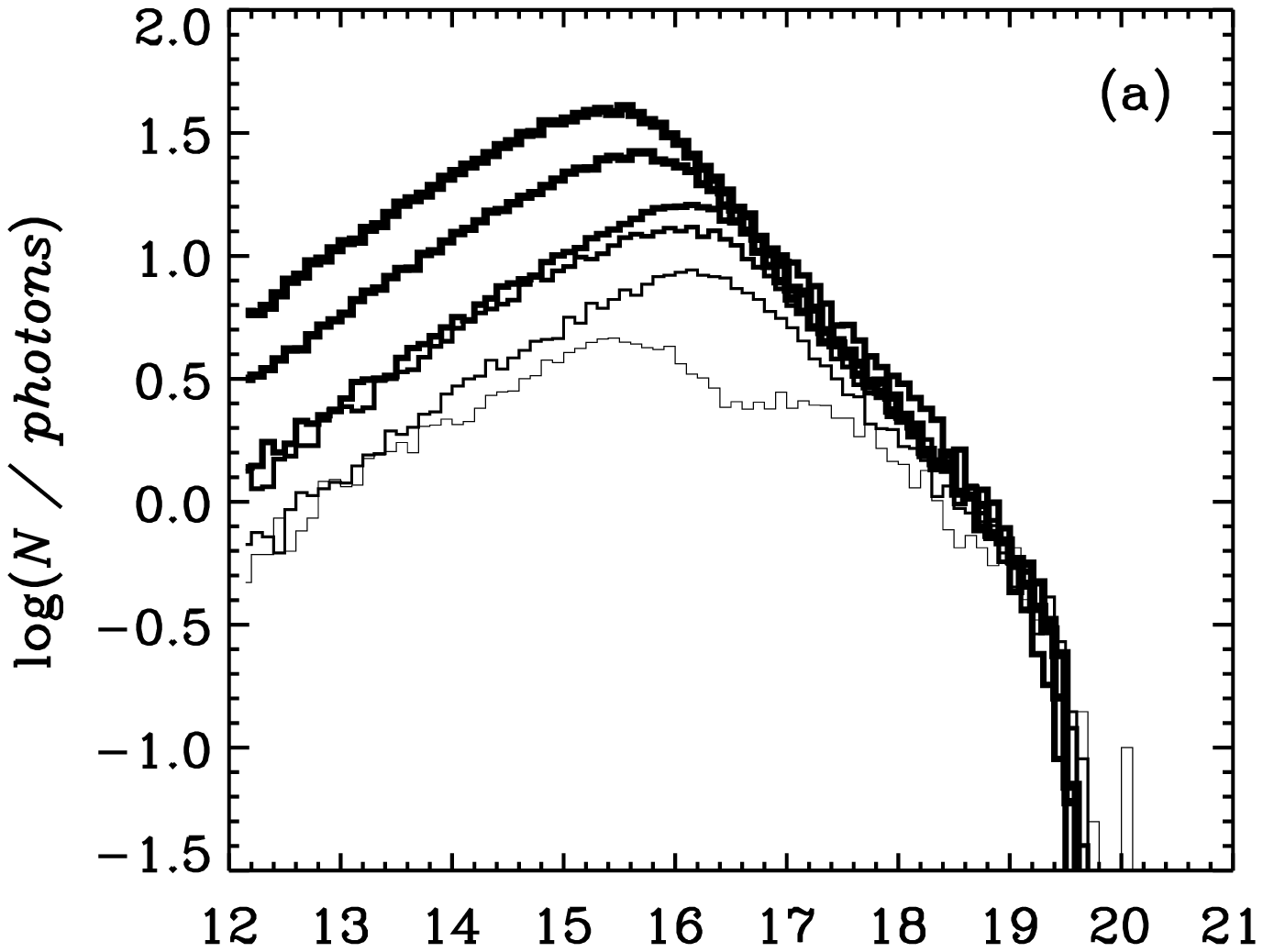}
\includegraphics{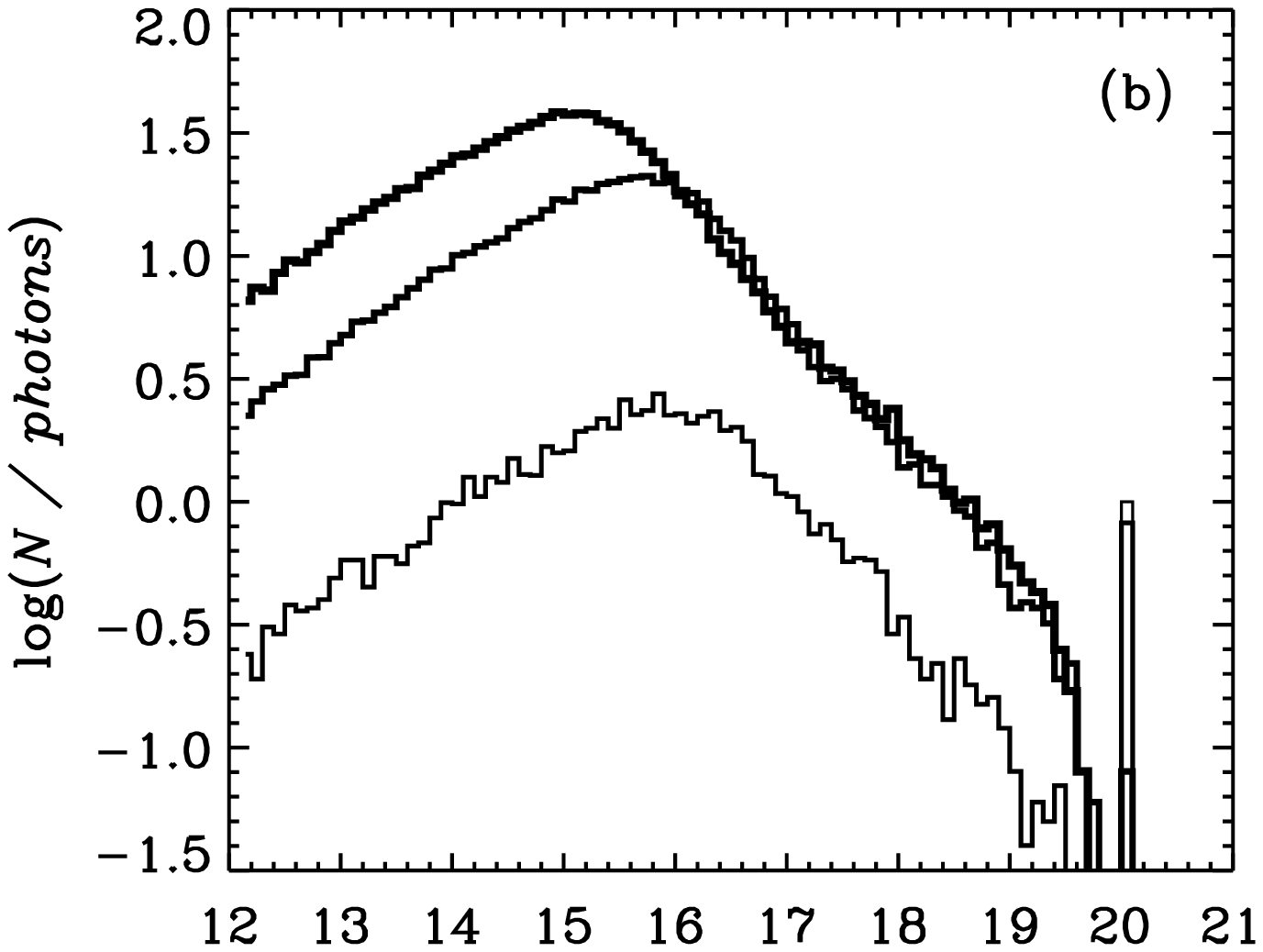}   
\includegraphics{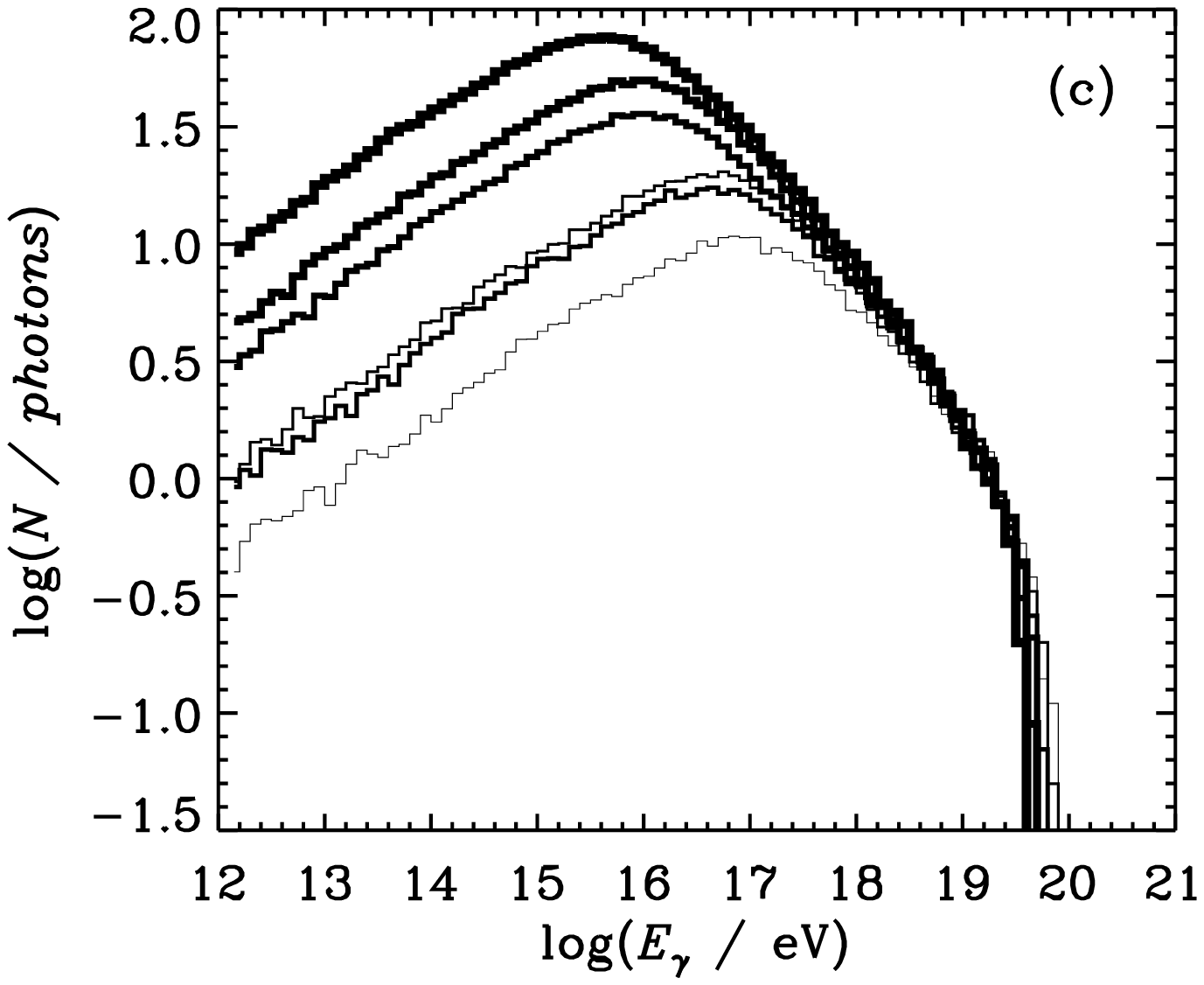}
\includegraphics{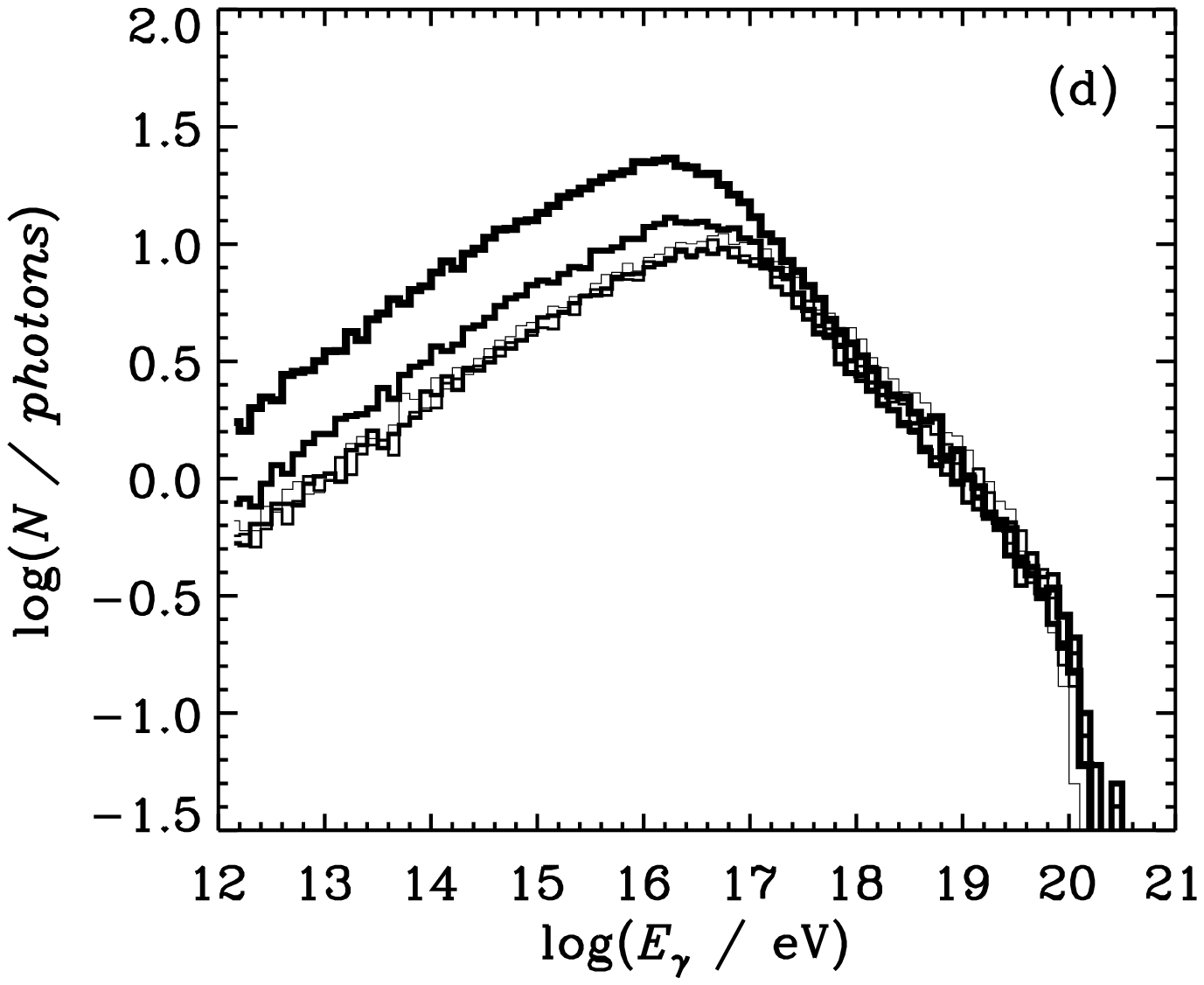}   
\caption[]{
Average numbers of secondary photons, per $\Delta 
log(E_\gamma) = 0.1$, produced in cascades initiated by primary 
photons with energies $10^{20}$eV (figure (a) and (b)) and 
$3\times 10^{20}$eV (figures (c) and (d)) in the magnetic field of 
the Earth for location of the Southern Auger Observatory. Specific
histograms show the results averaged over 
100 primary photons. They correspond to different arrival
directions defined by the azimuth angle, measured from the South Magnetic 
Pole, $\phi = 0^{\rm o}$ and the zenith angles  $Z = 75^{\rm o}, 60^{\rm o},
45^{\rm o}, 30^{\rm o}, 15^{\rm o}$, and  $0^{\rm o}$ (figures (a) and (c)
from the thickest to thinnest histograms), and $\phi = 180^{\rm o}$, and 
$Z = 75^{\rm o}, 60^{\rm o}, 45^{\rm o}, 30^{\rm o}$, and $15^{\rm o}$ (figures
(b) and (d)).}  
\label{fig4}  
\end{figure} 

\begin{figure} 
\vspace{12.truecm} 
\includegraphics{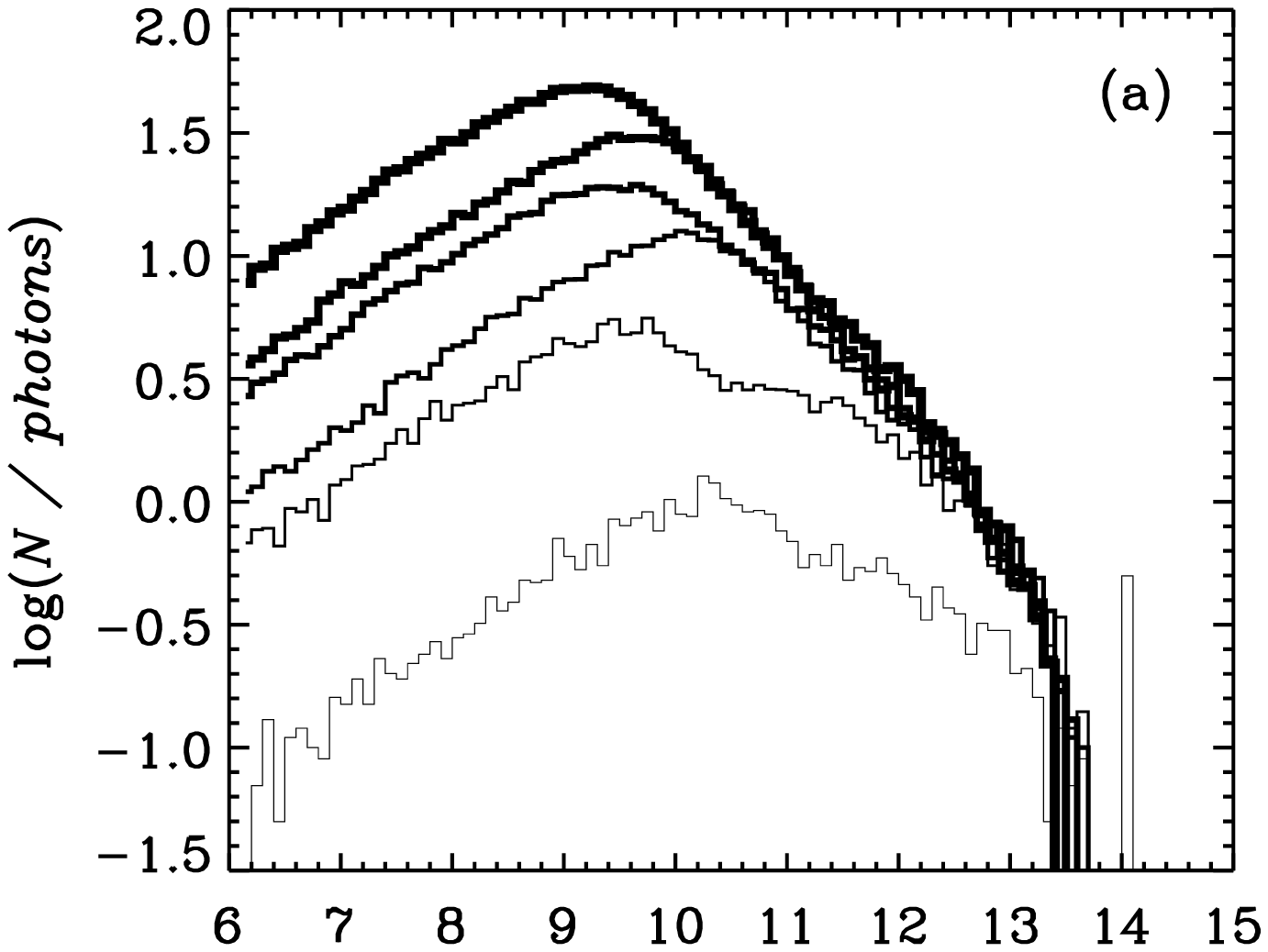}
\includegraphics{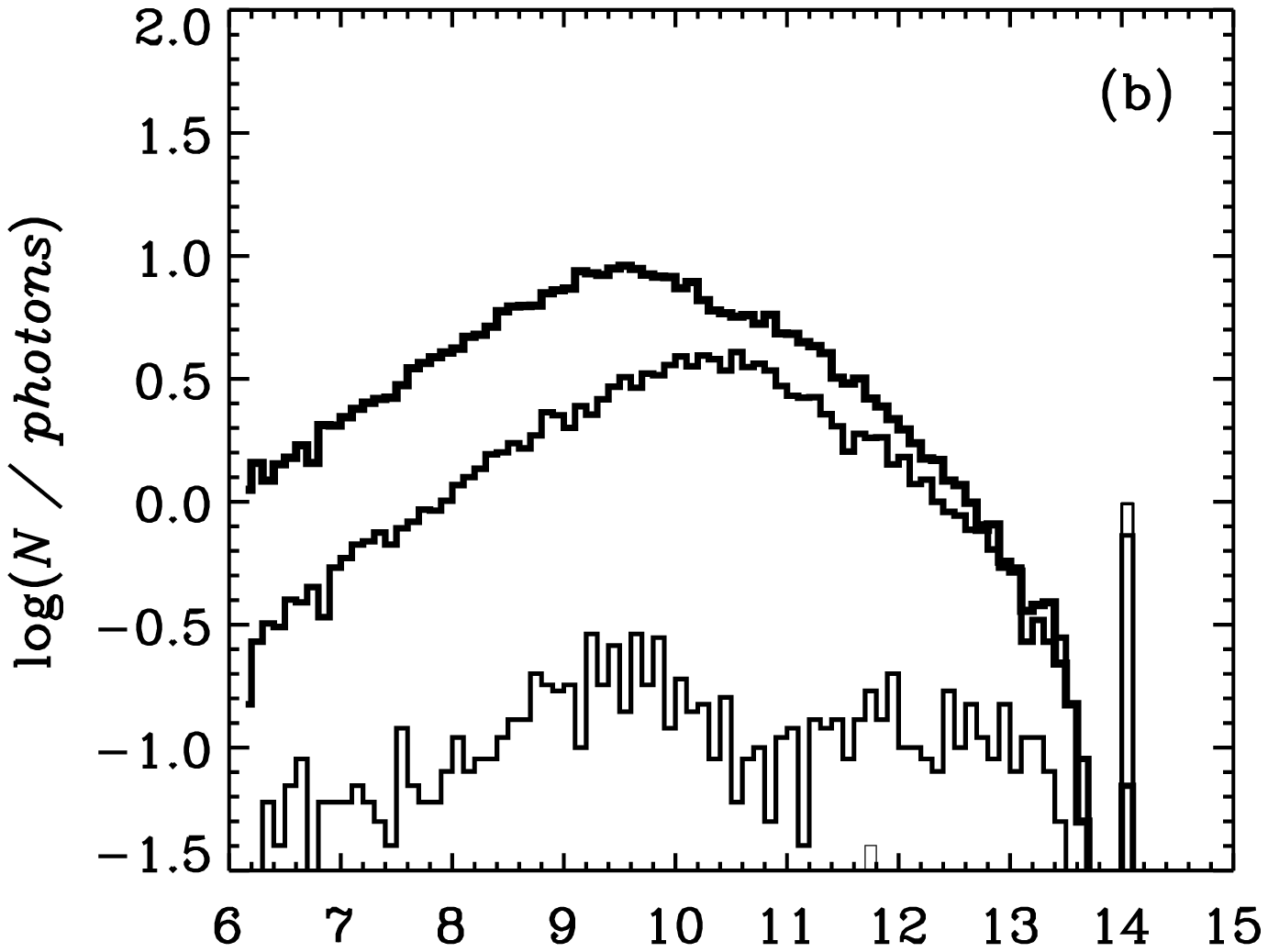}   
\includegraphics{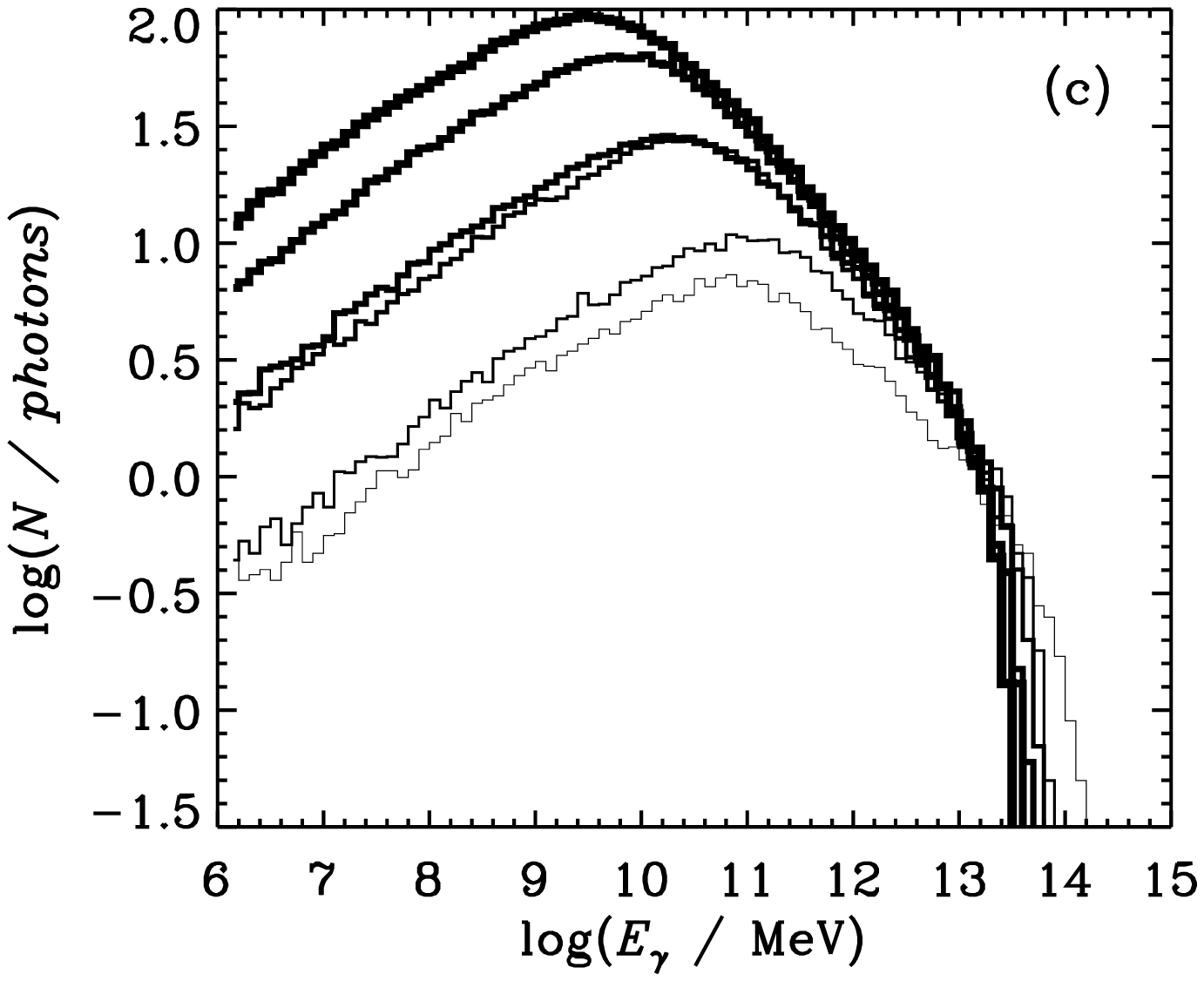}
\includegraphics{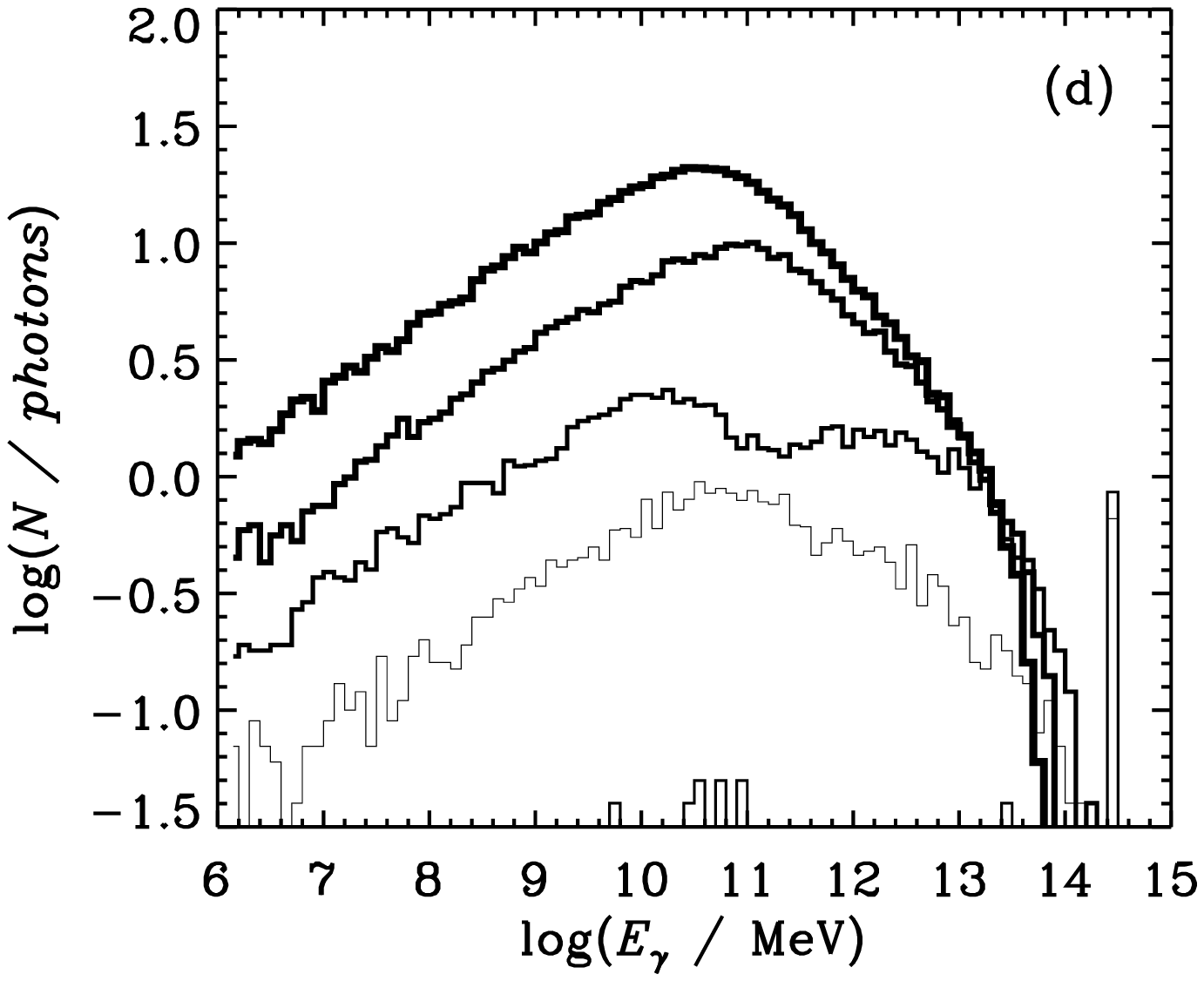}   
\caption[]{
As in Figs.~\ref{fig4} but for location of the Northern Auger
Observatory (also the location of the HiRes Fly's Eye and planned 
Telescope Array project).
The azimuth  is now measured from the North Magnetic Pole.} 
\label{fig5} 
\end{figure} 

\begin{figure} 
\vspace{10.truecm} 
\includegraphics{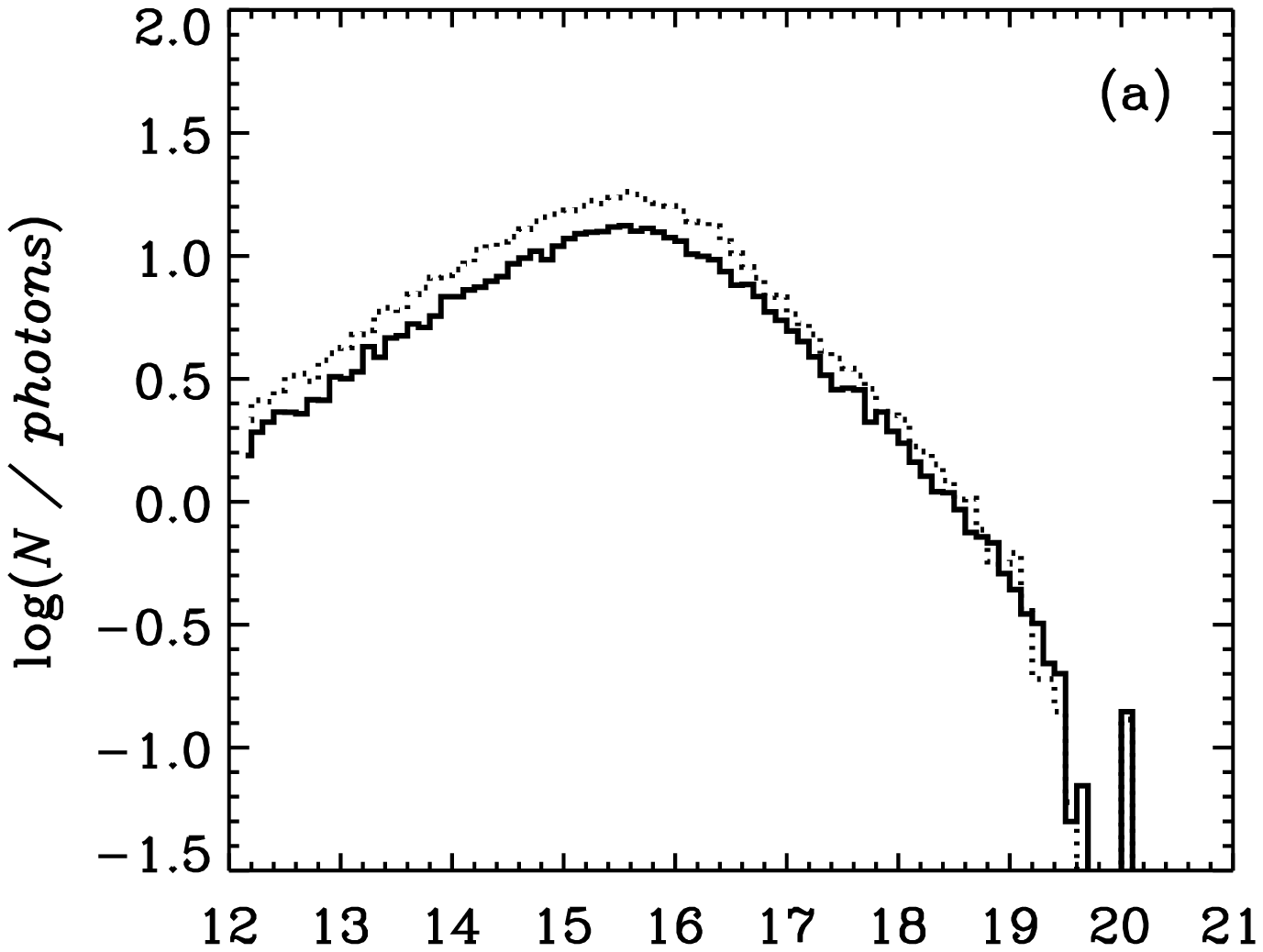}
\includegraphics{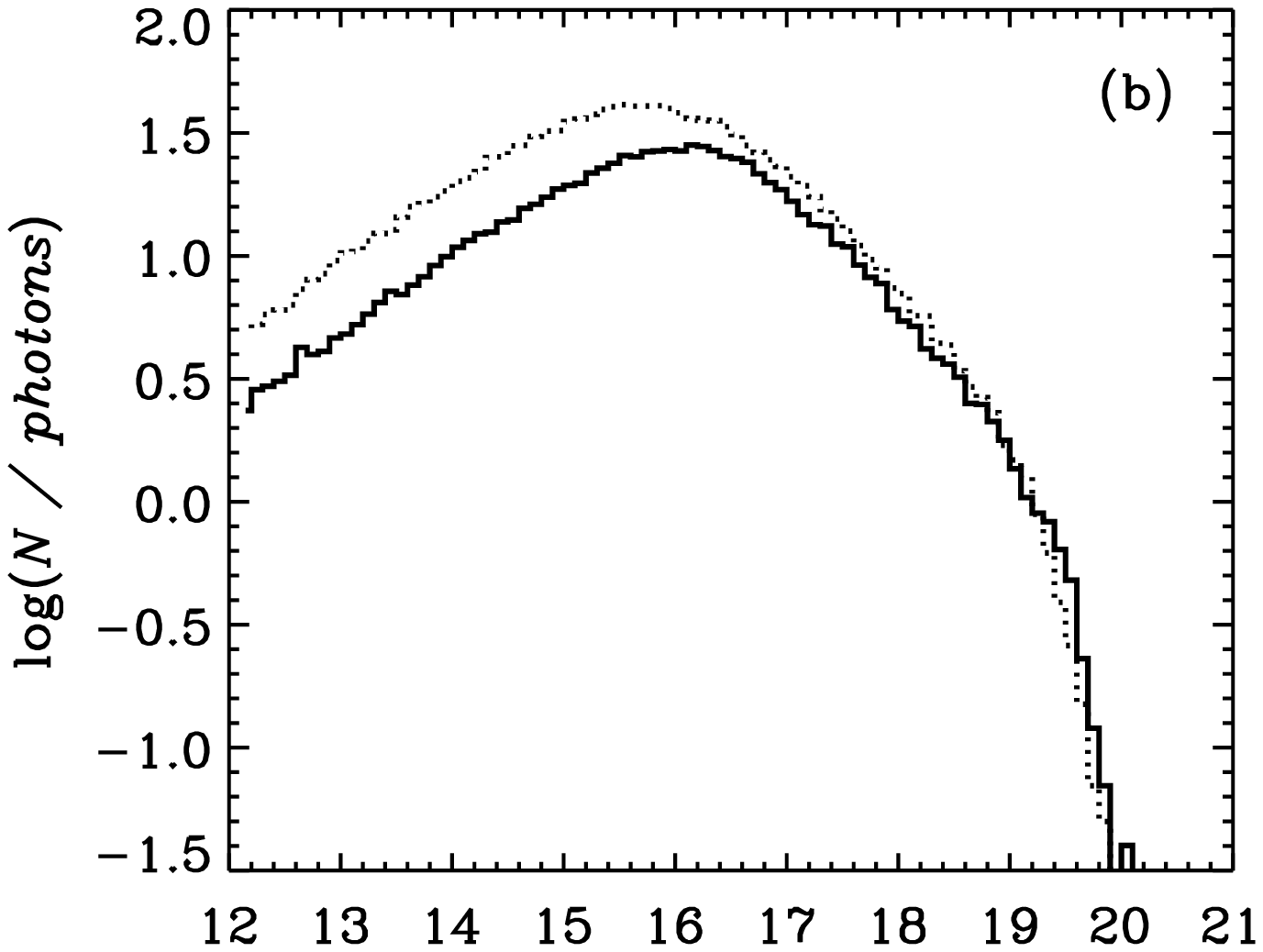}   
\caption[]{
The comparison of the number of secondary photons 
from cascades initiated by primary photons with energies $10^{20}$eV (a), 
and $3\times 10^{20}$eV (b), arriving to locations of the Southern (full 
histogram) and the Northern (dotted histogram) Auger Observatories. 
The results are averaged over 100 primary photons arriving randomly 
from the sky.} 
 \label{fig6} 
\end{figure} 

\begin{figure} 
\vspace{10.truecm} 
\includegraphics{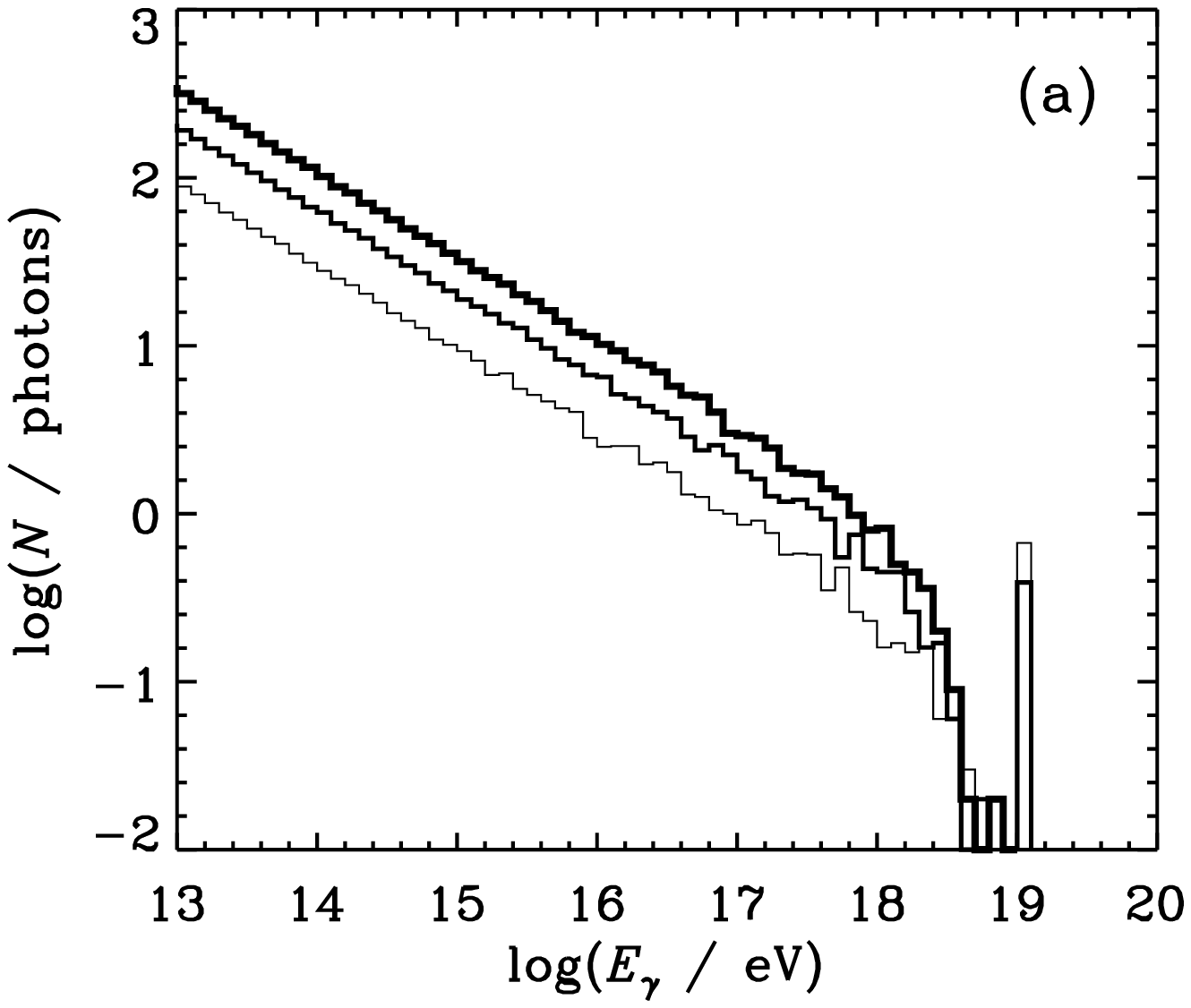}
\includegraphics{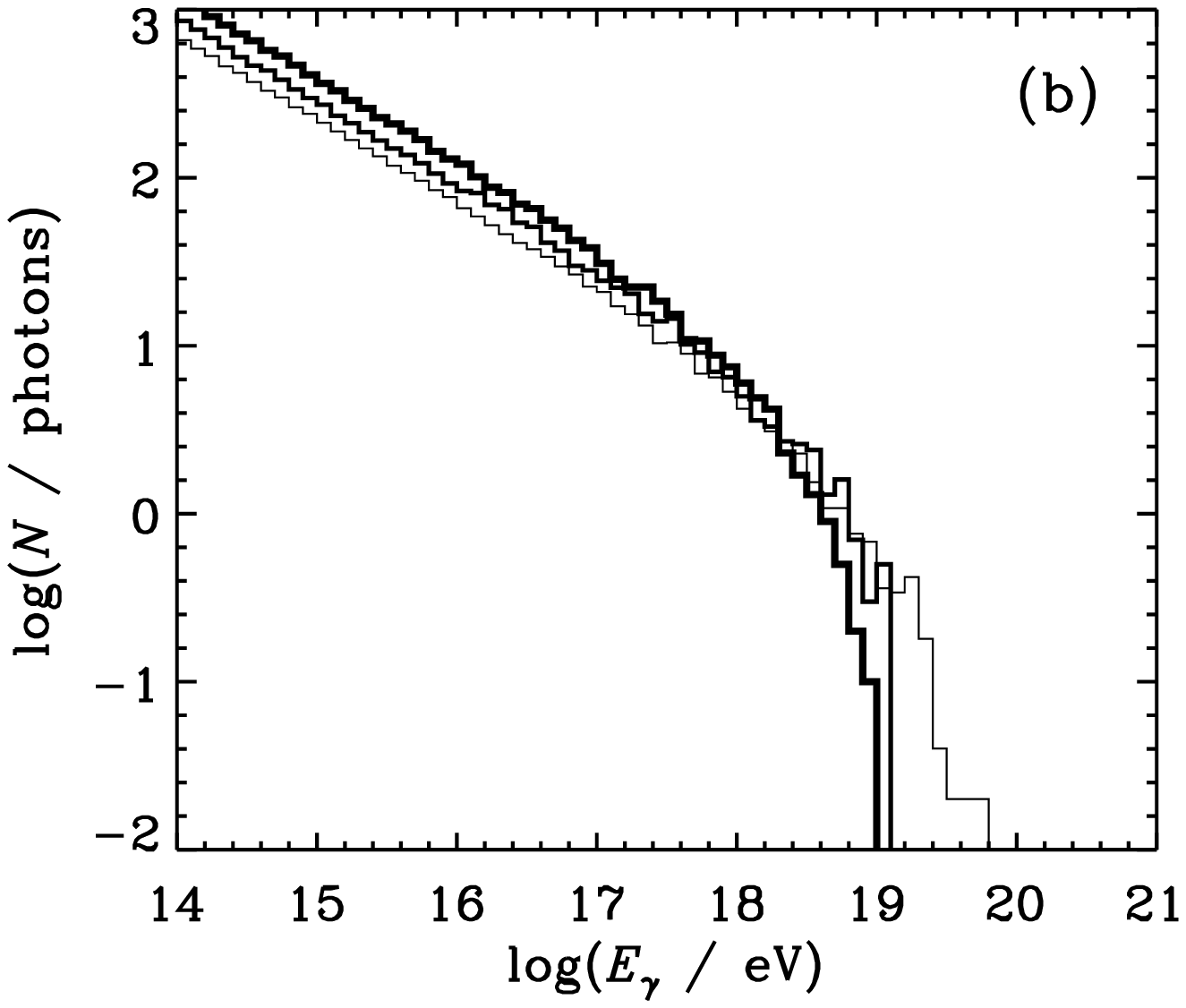}   
\caption[]{
Average number of secondary photons, 
per $\Delta$ log$E_\gamma = 0.1$, from cascades 
initiated by 100 primary photons with energies $10^{19}$ eV 
(figure a), and 10 photons with energies $10^{20}$ eV 
(figure b). Primary photons are injected randomly within a circle with 
the radius $s = 1.5r_{\rm s}$ around the Sun (the 
thickest full curve), $2r_{\rm s}$, and $3r_{\rm s}$ (the thinnest full 
curve). } 
\label{fig7}   
\end{figure} 

\begin{figure} 
\vspace{10.truecm} 
\includegraphics{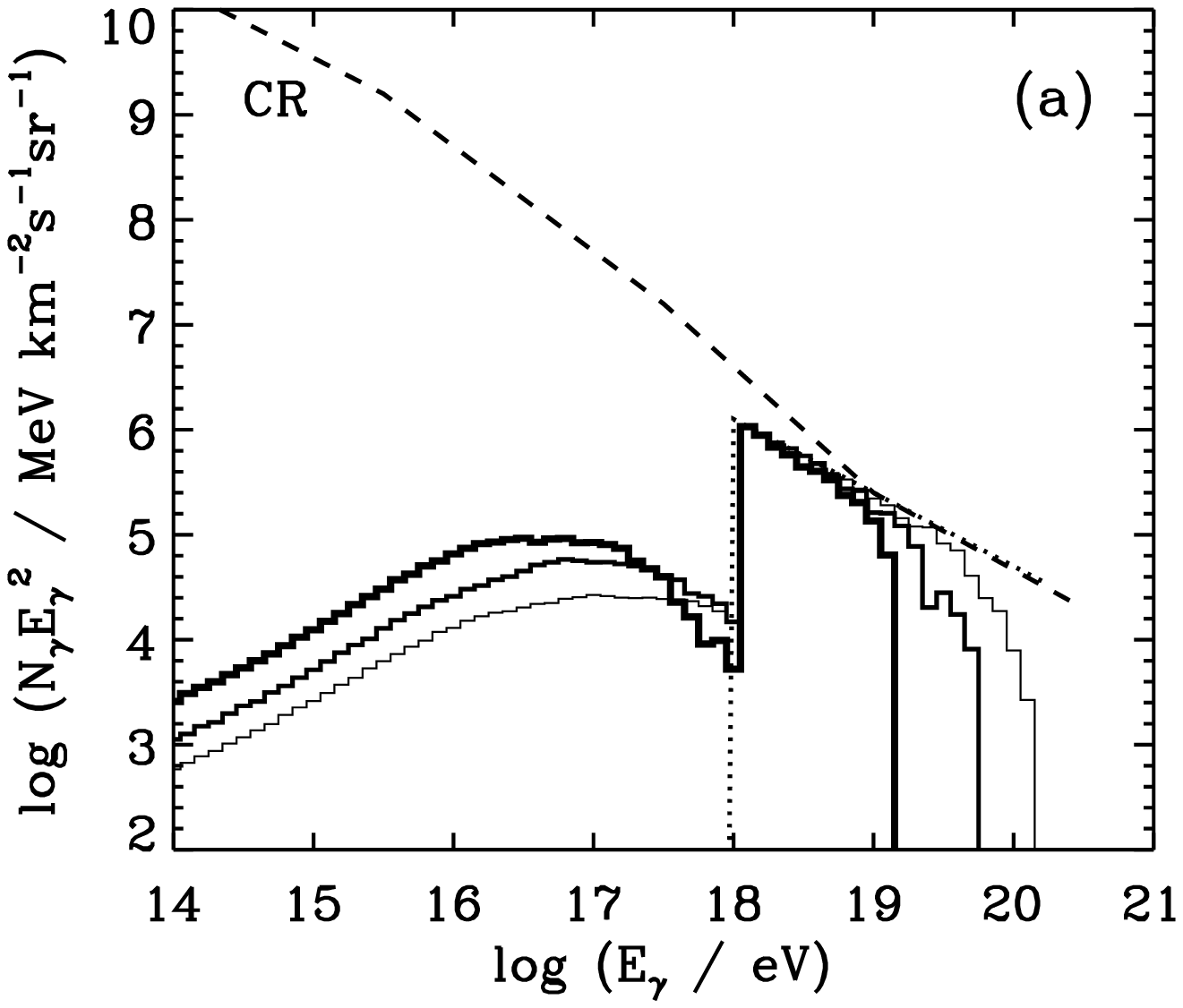}
\includegraphics{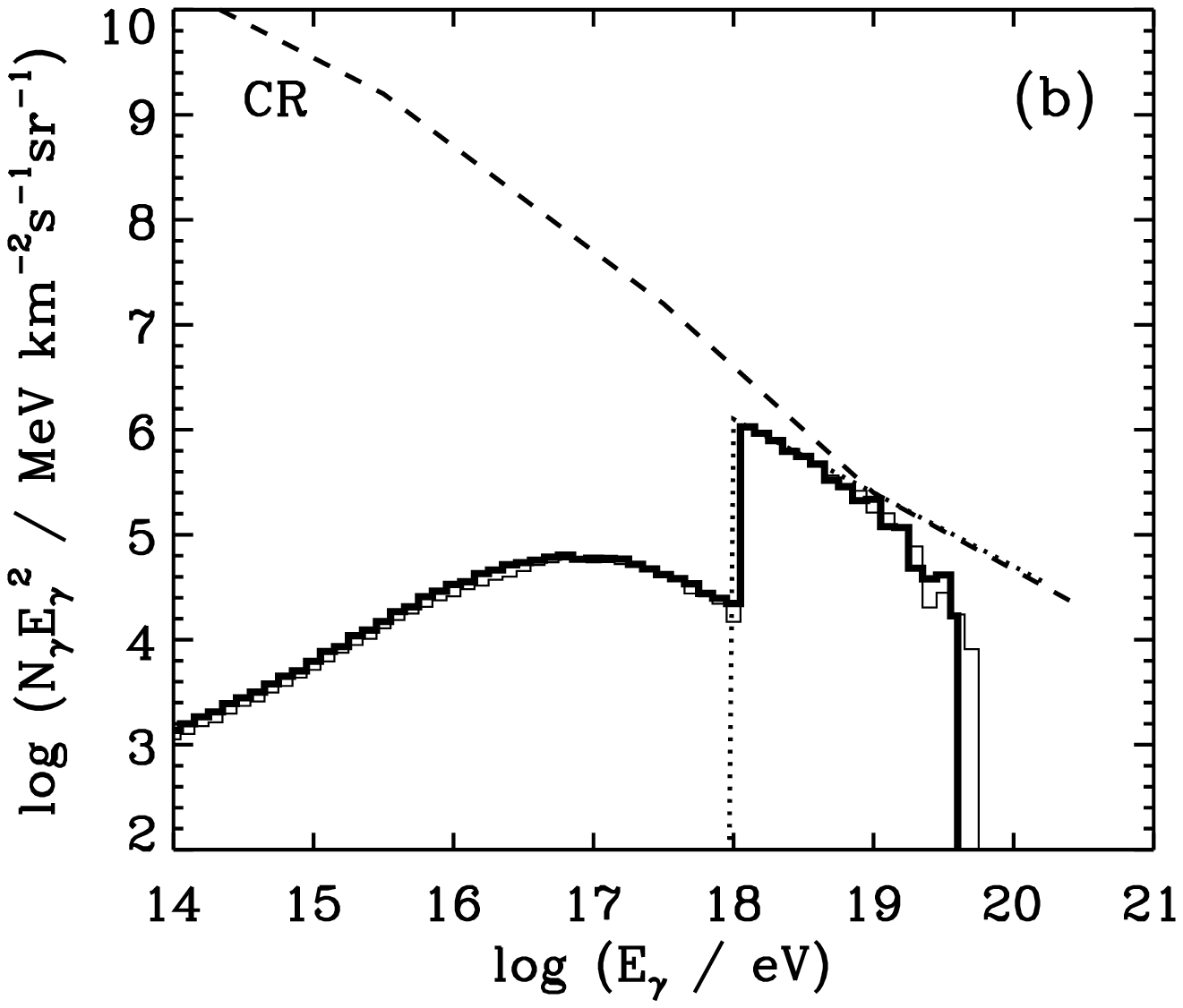}   
\caption[]{
(a) Spectra of secondary $\gamma$-rays 
(multiplied by the square of the photon energy) from cascades  
initiated by primary photons with the power law spectrum and
the spectral  index -2.7 
above $10^{18}$ eV and the cut-off at
$3\times 10^{20}$ eV (marked by the dotted curve). 
The spectra emerging from the Sun's magnetosphere are shown for
primary photons injected within a circle with the radius $s =
1.5r_{\rm s}$ around the Sun (the thickest full curve),  
$2r_{\rm s}$, and $3r_{\rm s}$ 
(the thinnest curve). (b) As in figure (a) but for the 
primary $\gamma$-ray spectrum injected within $s = 2r_{\rm s}$ and 
extending up to $3\times 10^{20}$ eV (thin curve) and 
$3\times 10^{21}$ eV (thick curve). The observed cosmic ray 
spectrum CR~\cite{yd98} is schematically marked by the 
dashed curve.} 
 \label{fig8} 
\end{figure} 

\begin{figure} 
\vspace{15.truecm} 
\includegraphics{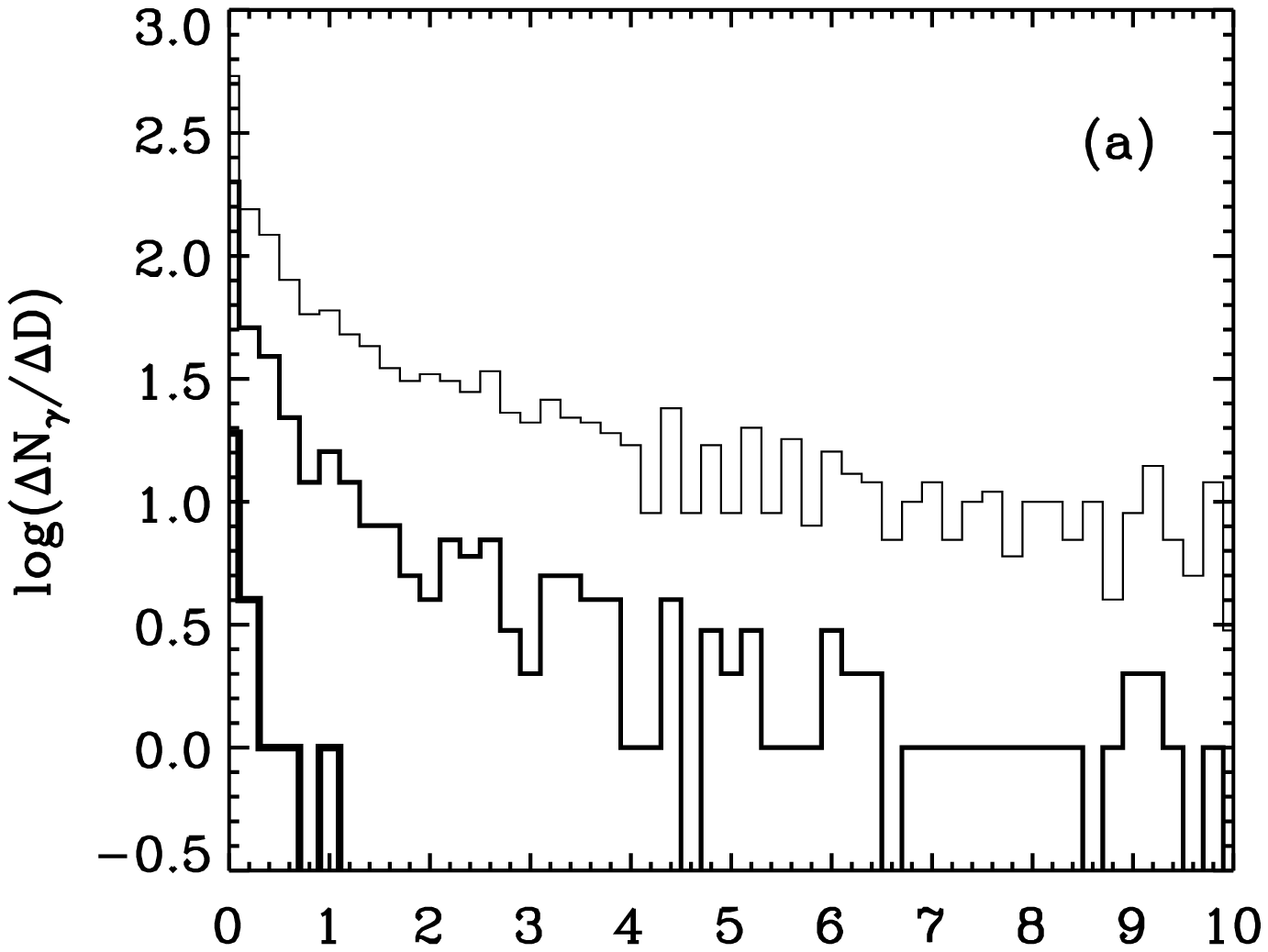}   
\includegraphics{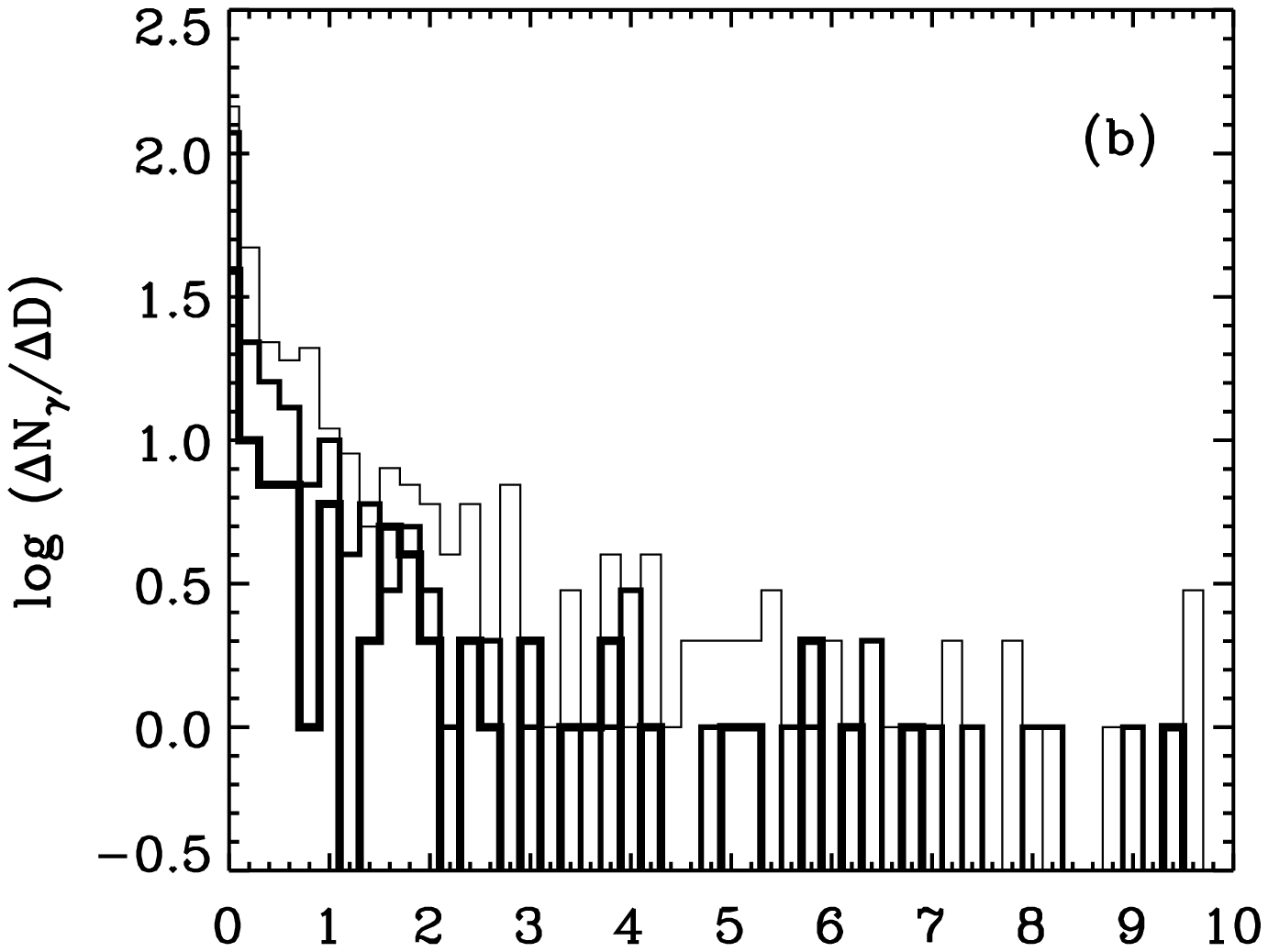}
\includegraphics{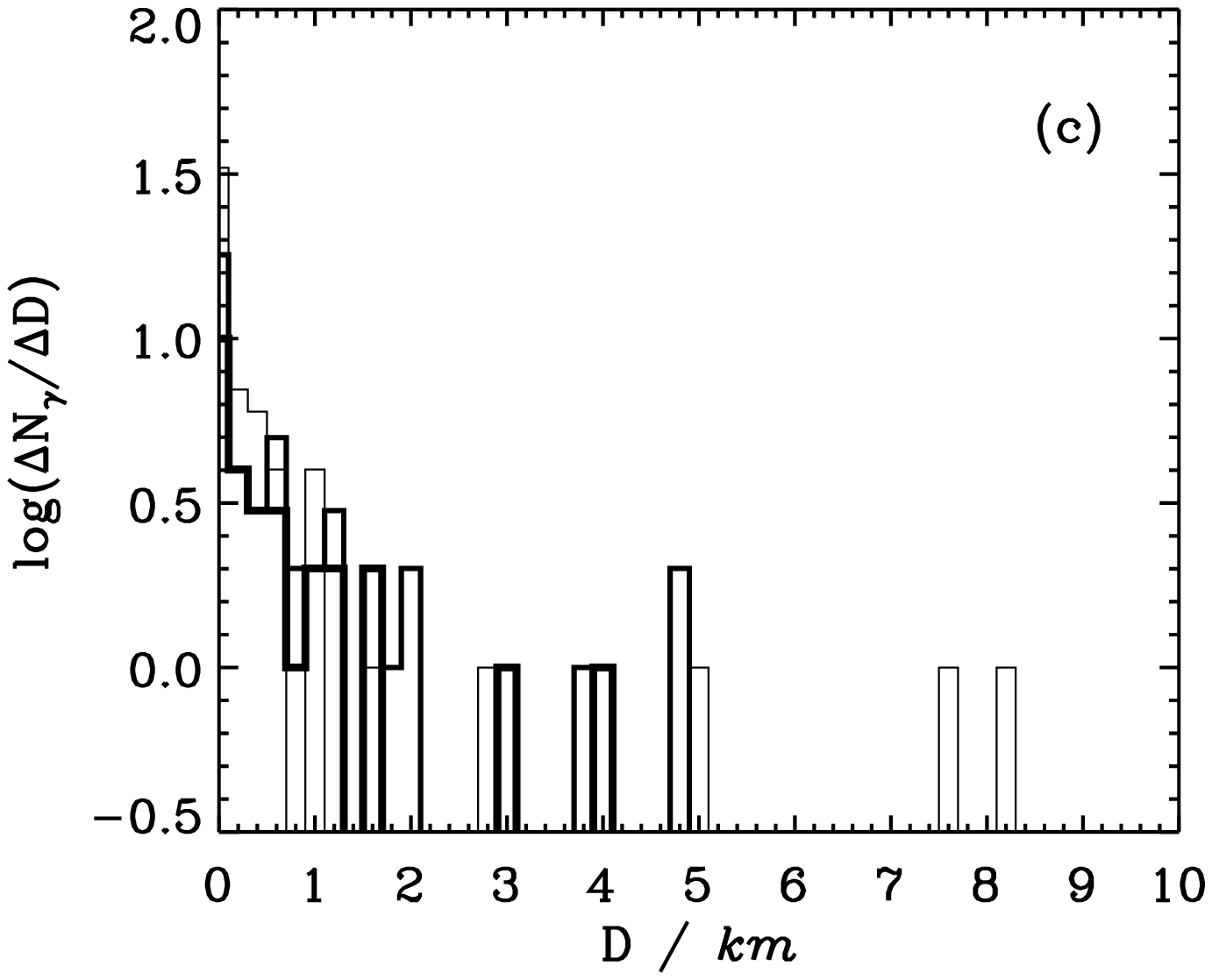}   
\caption[]{
Perpendicular extent of secondary photons, produced in the cascade 
initiated by primary photon in the Sun's magnetosphere, which fall on 
the Earth's atmosphere. Specific figures show the number of secondary 
photons $\Delta N_\gamma$ with energy above  $E_{\gamma ,s}$ in a ring 
with the width $\Delta D = 0.2$ km and radius $D$, centered on the 
direction of primary photon. 
The parameters of simulations presented in figures 
(a), (b), and (c) are given in the main text of the paper.
} 
\label{fig9} 
\end{figure} 

\end{document}